\documentclass[12pt,preprint]{aastex}

\shorttitle{Estimating the Ages of Bars}
\shortauthors{Gadotti \& de Souza}

\begin{document}

\title{The Vertical Stellar Kinematics in Face--On Barred Galaxies: Estimating the Ages of Bars}

\author{D. A. Gadotti\footnote{Present address: Laboratoire d'Astrophysique de Marseille,
2 Place Le Verrier, 13248 Marseille Cedex 04, France} \,and R. E. de Souza}
\affil{Departamento de Astronomia, Universidade de S\~ao Paulo, Rua do Mat\~ao, 1226,
05508-090, S\~ao Paulo-SP, Brasil}
\email{dimitri@astro.iag.usp.br,ronaldo@astro.iag.usp.br}

\begin{abstract}
In order to perform a detailed study of the stellar kinematics in the vertical axis of bars, we obtained
high signal--to--noise spectra along the major and minor axes of the bars in a sample of 14 face--on galaxies,
and used them to determine the line of sight stellar velocity distribution, parameterized as Gauss--Hermite
series. With these data, we developed a diagnostic tool that allows one to distinguish between recently
formed and evolved bars, as well as estimate their ages, assuming that bars form in vertically thin disks,
recognizable by low values for the vertical velocity dispersion $\sigma_z$.
Through $N$--body realizations of bar unstable
disk galaxies we could also check the time scales involved in the processes which give bars an important
vertical structure. We show that $\sigma_z$ in evolved bars is roughly around
100 Km/s, which translates to a height scale of about 1.4 Kpc, giving support to scenarios in which bulges
form through disk material. Furthermore, the bars in our numerical simulations have values for $\sigma_z$
generally smaller than 50 Km/s even after evolving for 2 Gyr, suggesting that a slow process is
responsible for making bars as vertically thick as we observe. We verify theoretically that the
Spitzer-Schwarzschild mechanism is quantitatively able to explain these observations if we assume that
giant molecular clouds are twice as much concentrated along the bar as in the remaining of the disk.
\end{abstract}


\section{Introduction}
In the last 10 years or so, bars have gradually become much more than just an intriguing dynamical
curiosity \citep[as in the pioneering studies of, e.g.,][]{too63,too64,kal72} to reveal its major role in the
formation and evolution of galaxies. Contributions to this change of perspective came from various
different kinds of analysis, and include the realization that
bars may induce the formation of spiral arms and rings \citep[e.g.,][see also
\citealt{but96}]{sch81,com85,but86}, an angular
momentum transfer to the outer parts of the galaxy, with consequences as an accumulation of gas
in the central regions \citep[e.g.,][]{ath92a,ath92b,fri93,fri95,sak99a,sak99b}, sweeping large scale chemical
abundance gradients \citep{mar94,zar94}, possibly building a reservoir of AGN fuel \citep{shl89,shl90},
and provoking central bursts of star formation \citep[e.g.,][]{ser65,ser67,car97,gad01}. Moreover,
a number of works came to develop a new bulge building scenario in which bars play a fundamental role
\citep[e.g.,][see also the recent review by \citealt{kor04}]{com81,kor82,kor83,des87,pfe90,com90,
bal94,kui95,nor96,cou96,pel96,ber98,mer99,bur99,ath99,bur04}.

In spite of the undisputed major relevance of bars in the evolution of galaxies, and even though one
of the major and basic concerns of any physical science is to measure time scales for natural phenomena,
we are not aware of any directed and systematic study on the ages of bars. Relevant points in which
such a work may have a substantial impact include the models of \citet{bou02}, in which bars
may be destroyed and rebuilt a few times in a Hubble period, the polemic and long sought correlation between
the presence of bars and AGN in galaxies \citep[e.g.,][]{mul97,ho97a,ho97b,kna00,lai02,cre03,lau04},
the debated frequency of bars at higher redshifts
\citep[and references therein; see also \citealt{elm04,jog04}]{vdb02,she03} and
obviously the formation history of galactic bulges.

Essentially, to estimate for how long a bar is evolving in a certain galaxy one has to measure its vertical
extent. This follows from the fact that, as was first shown by \citet{com81}, when bars form in disks they
are vertically thin, but the onset of vertical resonances rapidly makes bars grow thicker
in this direction. Another
possibility is that the hose instability \citep{too66,mer94}, that occurs whenever the velocity dispersion
in the vertical direction is 3 times smaller than in the plane of the disk,
also plays its role in this context, by raising
the vertical extent of the stellar orbits in the bar region. It is now generally agreed that these processes
are likely responsible for the existence of the so called boxy/peanut bulges \citep[see][]{burf99}. The
signature of bars' ages may thus be acquired from studying the kinematics along their vertical axis,
since it is by elevating the velocity dispersion in this direction that bars evolve and grow away from
the plane of the disk.

Another subject that belongs to this discussion regards the difficulties encountered, by what has been
until recently the standard
scenario for bar formation, in trying to explain the existence of bars in galaxies as early--type as lenticulars
\citep[see, e.g.,][and references therein]{gad03}. As the velocity dispersion of the stars in the disk rises,
it becomes more and more stable against bar formation through the disk instability. Moreover, the
presence of a conspicuous bulge also inhibits this instability \citep{too81,sel99}. However, \citet{ath03}
shows that the use of an unresponsive rigid halo in the older numerical experiments induced to a wrong
conclusion, namely, that a dark matter halo prevents the onset of the bar instability in the disk within.
In reality, the opposite is true: the exchange of angular momentum between disk and halo particles is able
to produce in fact even {\em stronger} bars, and might be a necessary ingredient to account for the
existence of barred lenticulars. A few other most important details that were not considered before have
been introduced by some authors and will be briefly discussed further on.

In this paper, we obtain suitable stellar kinematical parameters to develop a diagnostic tool that
enables us to estimate the ages of bars in a sample of 14 face--on galaxies. In \S\,2, we present
the sample and the observations done, while in \S\,3 we show how the kinematical parameters
were determined, and introduce our method for bar age estimates. Our results are presented
in \S\,4. To numerically assess the time scales involved in the vertical growth of bars, and in this
way get a deeper understanding on their evolution, several $N$--body simulations were performed.
These results are presented in \S\,5, along with a brief discussion concerning bar forming scenarios. In
\S\,6, we present a discussion on our results, also considering some of the possible implications on our
present knowledge about bar formation and evolution. Finally, \S\,7 summarizes this paper by
presenting our main conclusions. We used a value for the Hubble constant of
H$_0=70$ Km s$^{-1}$ Mpc$^{-1}$.

\section{Sample and Observations}

Relevant properties of the galaxies in our sample are shown in Table 1. While all of them are bright
face--on galaxies of the local universe, one can see that our sample
spans a variety of galaxy morphologies. Of the 14 objects there are 10 strongly barred
galaxies and 4 weakly barred ones. Moreover, 7 are of types S0 or S0/a, 3 are Sa and there are also
4 Sb galaxies. Also, our sample contains 4 galaxies with an identified companion that may be gravitationally
interacting and 5 with non--stellar nuclear activity. This variety is helpful in trying to evaluate
clues related, for instance, to the proeminence of the bulge, the bar strength and the gravitational
perturbation of a companion. The presence of galaxies with active nuclei might also be relevant to help in
understanding the role played by bars in the fueling of this phenomenon.

In Fig. 1 we show images of all the galaxies in our sample. Lines displayed horizontally help seeing to
what extent our spectra were taken along the bars' major and minor axes. For instance, one can see
that in the case of the SB0 galaxy NGC 4608 the spectra along the bar minor axis are all within the bar,
while those from the SBb galaxy NGC 5850 reach the region outside the bar, in the disk.

Our spectra were taken in two different sets of observing runs. One in the North, in the nights of
1999 May 7, and 2000 April 9 to 11, with the 2.3 m University of Arizona Steward Observatory
Bok telescope, on Kitt Peak; and the other in the South, in the nights of 2002 March 13 and December 1
to 5, with the 1.5 m European Southern Observatory telescope at La Silla. The instrumental set--ups
are, however, similar. In all runs we have used a Boller \& Chivens spectrograph, with a spatial resolution
of 0.8'' per pixel, a grating with a dispersion of 1 {\AA} per pixel, and an instrumental spectral
resolution of 1.1 {\AA}, giving a velocity resolution of 65 Km/s in the spectral region of the Mg I
feature at 5175 {\AA}, that is approximately the center of all spectra taken. The differences between the
North and South spectra are the slit width (2.5'' in the North and 2'' in the South), the average
seeing (1.5'' in the North and 1.8'' in the South), and the spectral range (typically 1000 {\AA}
in the North and 2000 {\AA} in the South).

The North spectra are composed from four 1800 s exposures with the slit oriented along the bar
major axis and two 1800 s exposures with the slit positioned along the bar minor axis. Whereas
in the South we have taken, respectively, four 2700 s exposures and two 2700 s exposures.
The slit was always centered in the galaxy nucleus. Thus,
the spectra along the bars' minor axis have generally a lower signal--to--noise ratio (S/N) than
those along the bars' major axis. This is also due to the fact that the surface brightness decreases
more quickly along bars' minor axis. Due to the difference in the telescope apertures the South spectra
have also a lower S/N than the North spectra, in general, even considering the larger exposure times
in the South runs.

We have also obtained spectra for several standard stars in every run. These are standards for
spectrophotometry, velocity measurements and to obtain Lick indices, spanning spectral types
from M to O. In the North we observed Feije 34 \citep[spectral type O,][]{oke90},
HR 3951 (G3V), 6458 (G0V), 6685 (F2I), 6770 (G8III), 6775 (F7V), 6806 (K2V), and HR 6868 (M1III),
HD 89449 (F6IV), 90861 (K2III), 92588 (K1IV), 136202 (F8III-IV), 155500 (K0III)
and HD 172401 (K0III). In the South the stars observed
were HR 1544 (A1V), 1996 (O9.5V), 3454 (B3V), 2429 (K1III), 2574 (K4III), 4267 (M5.5III), 4657
(F5V), 4995 (G6V), 5019 (G6V), and HR 5568 (K4V), HD 134439 (K0V), 37984 (K1III), 66141 (K2III),
71597 (K2III) and HD 92588 (K1IV). The spectral types are from \citet{hof91}.

All spectra were reduced and extracted from the spectral CCD images using the same standard
procedures with the {\sc onedspec} and {\sc twodspec.longslit} tasks from
{\sc iraf}\footnote{{\sc iraf} is distributed by the National Optical Astronomy Observatories,
which are operated by the Association of Universities for Research
in Astronomy, Inc., under cooperative agreement with the National
Science Foundation.}. Dark current, overscan and bias were treated as in imaging
\citep[e.g.,][]{mas97}, whereas
flatfielding required some extra care, that consisted in a response correction of the dome flatfields
to eliminate the continuum from the dome diffuse light, and an illumination correction with the
twilight sky flatfields \citep[see, e.g.,][]{mas92}.

To extract the spectra we used the {\sc iraf} {\sc kpnoslit.apall} task. To assure that there are
no relevant geometric distortions in our instrumental set--up, meaning that one single tracing
may be used to extract all spectra from the same spectral image, we have verified that the
dispersion axis is the same along the spatial axis, i.e., it is parallel in different positions along the slit.
For this we have observed the same standard star in 6 different positions along the slit in a single frame
and compared the tracing in all positions. The spectra were extracted for each galaxy along the major
and minor axes of the bars in the center and in 8 other different positions along the slit length.
To minimize the S/N drop
in the outer spectra these were obtained from a gradually larger spatial interval. These were centered at
$r=0''$, $r=2.05''$, $r=4.5''$, $r=11.9''$ and $r=19.3''$ at each side of the center of the galaxy along
each bar axis, where $r$ is the galactocentric radius. The full width of these bins are, respectively,
2.4, 4, 7.2, 15.2 and 21.6 arcseconds, approximately. Thus,
the spectra at $r=0''$, $r=4.5''$ and $r=19.3''$ are adjacent, i.e., there are no pixels between each
bin, but also they don't overlap, as well as those at
$r=2.0''$ and $r=11.9''$. But considering the seeing effects the only two pairs of independent spectra
are at $r=0''$ and $r=11.9''$, and at $r=2.0''$ and $r=19.3''$. As the S/N drops very quickly from the
center it was not possible to obtain spectra farther out.

The spurious contribution from the sky was determined with the light in the outskirts of the slit
(3' from the center in the North sample and 2' in the South), where the light contribution from the galaxy
is much smaller, and subtracted from the data.
Emission sky lines that were not eliminated in this step were manually corrected by direct interpolation.
This was especially necessary in the North sample due to the proximity of the city of Tucson
to the Kitt Peak \citep[see][where relevant sky lines over Kitt Peak are presented]{ken92}.
Cosmic rays and bad pixels were also removed with
statistical considerations. The extracted spectra were then continuum normalized and calibrated in
wavelength. The error in the latter step was verified to be around 10 Km/s in the region of interest
(i.e., at about the Mg I feature).

The next and final step was to bring all spectra to the local standard of rest (LSR).
This was of course first done with
the velocity standard stars [with velocities available in \citet{abt72} and in the Astronomical
Almanac], whose corrected spectra were then used to bring the other ones to the LSR. This was done with
the cross correlation technique \citep{ton79} and the fact that we had spectra available from stars
of many different spectral types was helpful in minimizing the errors caused by template mismatch.
As the S/N per pixel is an essential parameter in calculating the errors in the derived line of sight velocity
distributions (LOSVDs), they were determined for every galaxy spectrum.
We found that S/N $\sim40-50$ in the central regions of the galaxies while it drops to S/N $\sim10-20$
in the outermost spectra.

In Fig. 2 we show some illustrative examples of the spectra obtained. The two upper panels refer to
spectra obtained along the bar major axis of NGC 4608 and NGC 4579, both from the North sample.
The lower panel shows spectra obtained along the minor axis of the bar in NGC 1387, from the South
sample. It is worth noticing how the higher values for $\sigma_z$ in the latter produce a larger width
in, e.g., the Mg I lines, and the H$\beta$ and [O III] emission lines in the  LINER/Sey 1.9 galaxy
NGC 4579, that are particularly strong in the center, as expected.

\section{Kinematical Parameters and Bar Age Estimates}

\subsection{Determining the LOSVDs}

The method we have chosen to determine the LOSVDs with our data is the line profile fitting
in the pixel space through Gauss--Hermite series \citep{vdm93}. In this case, assuming,
as is generally done, that the main difference between the spectrum of a galaxy and that
of a suitable template star is due to the stellar velocities in the galaxy \citep[see, e.g.,][]{bin98},
following a distribution close to a gaussian, one may write the line profile in the galaxy spectrum as
a function of the line of sight stellar velocity $v$, as:

\begin{equation}
L(v)=\frac{\gamma \alpha(w)}{\sigma} \sum_{j=0}^4 h_j H_j(w),
\end{equation}

\noindent where

\begin{equation}
\alpha(w)=\frac{1}{\sqrt{2 \pi}} e^{-w^2/2},
\end{equation}

\noindent and

\begin{equation}
w\equiv \frac{v-v_0}{\sigma}.
\end{equation}

\noindent In these equations, $\gamma$ is the parameter that adjusts the line depth, $\sigma$
is the stellar
velocity dispersion (in our case, $\sigma\approx\sigma_z$), $v_0$ is the average radial velocity of the
system (in our case, $v_0\approx0$), $h_j$ are numerical
constants, and $H_j(w)$ are the orthogonal Hermite polynomials
\citep[see][]{abr65}. We may rewrite Eq. (1) as:

\begin{equation}
L(v)=\frac{\gamma \alpha(w)}{\sigma} [1 + h_3 H_3(w) + h_4 H_4(w)],
\end{equation}

\noindent where

\begin{equation}
H_3(w)=\frac{1}{\sqrt{6}}(2\sqrt{2}w^3-3\sqrt{2}w),
\end{equation}

\noindent and

\begin{equation}
H_4(w)=\frac{1}{\sqrt{24}}(4w^4-12w^2+3),
\end{equation}

\noindent as we have truncated the Gauss--Hermite series in the terms of order 4 (higher order
terms are not retrieved reliably, in general), and since $h_0=H_0(w)=1$ e $h_1=h_2=0$
\citep[see][]{vdm93}.

Thus,
to derive the kinematical parameters along the vertical axis in the major and minor axes of the
bars of the galaxies in our sample, we have developed an algorithm that may use template
spectra of up to 5 different stars to determine from the galaxy spectrum
$\gamma$, $v_0$, $\sigma$, $h_3$ and $h_4$. However, to optimize efficiency we have used up to 3
different template stars, verifying that in our case adding more stars would not result in a
significant improvement on the quality of the results. In this way, errors from template
mismatch are severely reduced \citep[see, e.g.,][]{rix92},
as our code is also able to determine the contribution of each stellar type
to the galaxy spectrum that maximizes the quality of the fit (i.e., minimizes the $\chi^2$ value).
As shown by \citet{vdm93} this method here employed minimizes the correlation between
the errors in the different kinematical parameters that determine the line profile, which makes
the error evaluation safer. Moreover, these authors also show that it is less sensible to
template mismatch, as the line profile may fit into the small differences of the spectral
properties of the template stars and the galaxy, through adjusts in the higher order moments
of the Gauss--Hermite series. Hence, the relevant kinematical parameters we have determined
for each galaxy spectrum obtained are $\sigma_z$ and the third and fourth order moments
of the LOSVD parameterized as Gauss--Hermite series, $h_3$ (the skewness
of the velocity distribution) and $h_4$ (its kurtosis). This was done
twice for each spectrum, since, to check the consistency of our results, we have also
parameterized the LOSVDs with pure gaussians, i.e., with $h_3=h_4=0$.

Figure 3 illustrates how $h_3$ and $h_4$ modify the shape of the distribution. As can be
seen, $h_3$ is responsible for asymmetric deviations. A negative value for $h_3$ means
that there is an excess of stars whose velocities are lower than the average system
velocity, while the opposite is of course true for a positive $h_3$. This is why an anticorrelation
between $h_3$ and $v$ is generally found in edge-on galaxies to be a signature of a cold,
rapidly rotating system \citep{fis97,chu04}. On the other hand, a non--negligible $h_4$
introduces symmetric deviations. A negative value for $h_4$ indicates a higher number of
stars with line of sight velocities close to the average velocity, turning the LOSVD pointy, whereas
a positive value is a sign that the distribution is wider near the average velocity. Note also that
typically $h_3$ and $h_4$ are very small and in the range $[-0.1,0.1]$ and that to retrieve reliable values
for these parameters the S/N in the spectrum must be higher than about 50. This means
that a meaningful discussion on these deviations in our work can only be done when regarding
the central spectra. Even so, it is evident, however, that using a generalized gaussian
(Gauss--Hermite series) to
parameterize the LOSVDs maximizes the quality of the fit and the reliability on the values
of the kinematical parameters.

To certify that our code is able to produce reliable estimates a series of tests was performed.
In these tests, a stellar spectrum was artificially red shifted and widened by a known LOSVD,
resulting in an artificial galaxy spectrum, having also a spectral resolution of 1 {\AA} and a
S/N equal to 30, to match the typical characteristics of the spectra in our sample. For the tests
we chose generally K giant stars whose spectra were taken from the Elodie archive,
that contains high resolution spectra for many stars \citep[see][]{pru01}.
We then used our algorithm to retrieve
the LOSVD from the synthetic galaxy spectrum using as template the stars for which we have
obtained the spectra. The results have been always excellent: the solution found by our code, i.e.,
the template spectrum dislocated and widened by the determined LOSVD has always been
very similar to the original synthetic galaxy spectrum. In any case, when determining every LOSVD
in this work we have carefully verified that the solution found is indeed similar to the real galaxy
spectrum. Figure 4 shows a clarifying example of the fits produced by our code to the spectrum
of NGC 1302 at 2.0'' from the center along its bar major axis. Note, moreover, that, to avoid
infinite values for $v$, our code truncates it, following the relation
$\langle v_e^2\rangle=4\langle v^2\rangle$, where $v_e$ is the escape velocity in virialized
systems \citep[see][]{bin87}.

It is worth noticing how we have selected the template stars in the running of our code to
determine the LOSVDs. We followed a careful criterion to minimize even further errors
from template mismatch. For every galaxy spectrum we first ran the code using a single
stellar template spectrum. This was done with all stars observed under the same
instrumental setup. The star that provides the best fit, i.e., the lower $\chi^2$, is the
first star selected. After that, we ran the code again using as template spectrum a combination
of the spectra from the first selected star and all of the other remaining stars separately.
Initially, this combination is an average of both stellar spectra but the algorithm is able to
give weights to each spectra in order to achieve better results. Again, the pair of stars
that produces the best fit is selected. Finally, the algorithm is applied once again, using the
spectra of the stellar pair selected combined with the spectrum of each of the remaining stars,
to determine the third suitable template star in the same fashion. Generally, with three stars we
have achieved the best $\chi^2$ but, in some cases, using only two stars produced better results.
Only one set of template stars was used to retrieve the LOSVDs from all spectra of a given galaxy,
since we have verified that the best set of stars does not change when running the code for
spectra taken at different galactocentric distances.

Another cause of uncertainties from template mismatch regards the chemical abundances of the template
stars. Although most of the template stars we use are K giant or similar,
their metallicity (the [Fe/H] ratio)
compared to the solar one ranges from $-0.62$ (HR 6775) to $+0.25$ (HR 3951) in the North and
$-1.92$ (HD 134439) to $+0.07$ (HR 2429) in the South, considering the estimates in \citet{str01}.
Although we did not find the necessary data in the literature this likely means that our
template stars span also a significant range in the [Mg/Fe] abundance ratio. This obviously plays a role
in studies like ours where Mg and Fe lines are used to calculate the kinematical parameters.
This should be suitable to account for differences in the abundance of the galaxies and the stellar
templates. Interestingly, following our scheme to choose the most suitable template spectra, stars
with [Fe/H] values too different from solar were not considered suitable by our code.
For instance, in the North sample the two
most chosen stars were HR 6806 ([Fe/H] $\approx -0.30$) and HR 3951, and in the South sample these
were HD 37984 ([Fe/H] $\approx -0.55$) and HR 2574 ([Fe/H] $\approx -0.16$), with no significant or
systematic differences to galaxies that are early--type or late--type. In fact, one should expect
such a result as our measurements concern mostly the central regions of galaxies whose morphological
types are also mostly early.

The spectral region used to measure the kinematical parameters of the galaxies in our sample is the one
that contains mostly absorption lines from the photosphere of, typically, K giant stars, in the range
from, approximately, the Mg I triplet at $\lambda=5175$ {\AA} to the Na I feature at $\lambda=5893$ {\AA}.
This region also includes relevant lines in this respect, as the Fe I + Ca I lines at $\lambda=5265$ {\AA} and
the Fe lines at $\lambda=5328$ {\AA}. In a significant part of the cases studied here a narrower spectral
range was used, that excludes the Na I feature. The H$\alpha$ and [H$\beta$] lines, at, respectively
$\lambda=6563$ {\AA} and $\lambda=4861$ {\AA}, were excluded from the analysis as to avoid
spurious results from the complicated emission from gas. In the cases of galaxies with active nuclei,
emission lines like, e.g., [O III] ($\lambda=5007$ {\AA}) were automatically excluded from the analysis
by our code, that is able to ignore lines that are too discrepant in the galaxy and template spectra.
Nonetheless, we mention a potential source of uncertainty in the spectra of galaxies that
present a strong [N I] emission line (the doublet at $\lambda\lambda5198,5200$ {\AA}), such
as NGC 4579 (see Fig. 2). As shown by \citet{gou96} this feature may affect line strength measurements of
the Mg I triplet. Three other galaxies in our sample show also this [N I] emission line: NGC 2665, 4984 and
NGC 5701. Although LINERs, the spectra of NGC 1326 and NGC 4314 present only a small feature,
while in that of NGC 4394 we have not detected the line. Inspecting our results we find no clear systematic
trends regarding this issue.

\subsection{The Ages of Bars}

Having measured the value for $\sigma_z$ along the major and minor axes of the bars in our
galaxy sample, we are ready now to, as in a first order approach, distinguish recently formed
and evolved bars. But before going on, let us discuss what one expects to find. We argue that
to make this distinction one has to rely not only on the value of $\sigma_z$ in the bar, but also on its
radial behavior in both axes, since the contribution from the bulge and the disk must also be considered.
Being a stellar system supported by pressure, bulges will, in general, contribute to a rise in the
stellar velocity dispersion in the center, irrespective of the bar age. On the other hand, since
in some cases the outer spectra along the bar minor axis reach areas where we are essentially
measuring the value for $\sigma_z$ without relevant contributions from the bar and the bulge, i.e.,
the vertical velocity dispersion in the disk, in these cases, a comparison between the values
of $\sigma_z$ in the bar and in the disk is most valuable. It should also be kept in mind that
a significant variation in the velocity dispersion from galaxy to galaxy (even with similar
morphological types) is expected.

\citet[see also \citealt{bin98}]{del65} estimates that $\sigma_z\sim15$ Km/s in the solar neighborhood
for K giant stars. In general, similar values are found for other spectral types. The velocity dispersion
in the radial, $\sigma_r$, and azimuthal, $\sigma_\varphi$, directions are, respectively,
$\sim$ 30 Km/s and $\sim$ 20 Km/s. In fact, in the Galactic disk, typically, $\sigma_z/\sigma_r\sim0.5$
and $\sigma_{\varphi}/\sigma_r\sim0.6$, and thus $\sigma_r>\sigma_{\varphi}>\sigma_z$. As the bar length
in the Galaxy is estimated to be around 3 Kpc \citep{bli91,mer04}, and also considering
the study of stellar orbits in disk potentials \citep{bin87}, these values are
expected to be representative for the
outer disk in late--type spiral galaxies (Sb--Sc). As the velocity dispersion in the disk rises inward
\citep{vdk81}, at a galactocentric distance of about the disk length scale $h$,
which is 3.5$\pm$0.5 Kpc in the Galaxy, $\sigma_z\sim30$ Km/s. Typical values for $\sigma_z$ in the center
of this class of galaxies are around 50 Km/s \citep{bot93}. For lenticular galaxies,
$\sigma_z\sim100$ Km/s in the center, including the contribution from the bulge \citep{mce95}.
Farther out $\sigma_z\sim50$ Km/s in lenticulars \citep{fis97}.

We can also make an evaluation of the relation between $\sigma_z$ and the height scale in the disk.
In the epicycle approximation \citep{bin87}, the vertical oscillation of the stars in the disk in cylindrical
coordinates ($r$,$\varphi$,$z$) is given by $\ddot{z}=-\nu^2z$, where $\nu$ is the epicycle vertical
frequency, given by:

\begin{equation}
\nu^2=\left(\frac{\partial^2\Phi}{\partial z^2}\right)_{(r,z=0)},
\end{equation}

\noindent where $\Phi$ is the disk gravitational potential. Also, in a highly flattened system,

\begin{equation}
\frac{\partial^2\Phi}{\partial z^2}=4\pi G\rho(r,z).
\end{equation}

\noindent And thus,

\begin{equation}
\nu^2=4\pi G\rho(r,z=0).
\end{equation}

\noindent This means that $\nu$ depends only on the mass density in the plane of the disk, and
that in a recently formed bar, assuming that bars are a global dynamical disk instability, the
stars are oscillating with this vertical frequency. For the Galaxy, in the solar neighborhood,
$\nu=(3.2\pm0.5)\times 10^{-15}$ s$^{-1}$.

On the other hand, through the reasonable assumption that $\nu$ is independent of $z$, we
have:

\begin{equation}
z=z_0{\rm sin}(\nu t+\phi_0),
\end{equation}

\noindent where $z_0$ is the disk height scale, $t$ is time, and $\phi_0$ is a phase constant.
This implies in:

\begin{equation}
v_z=\dot{z}=z_0\nu {\rm cos}(\nu t+\phi_0),
\end{equation}

\noindent and

\begin{equation}
\langle v_z^2 \rangle = \sigma_z^2 = \frac{1}{2}z_0^2\nu^2.
\end{equation}

\noindent Thus, considering the disk of the Galaxy typical,

\begin{equation}
z_0=\frac{\sqrt{2}\sigma_z}{\nu}=215 \ (\sigma_z/15 \ {\rm Km} \ {\rm s}^{-1}) \ {\rm pc}.
\end{equation}

\citet{edv93} show that for the thin disk of the Galaxy $\sigma_z=18$ Km/s while $\sigma_z=39$ Km/s
for the thick disk. Following our Eq. (13), this means that $z_0$ is, respectively, 258 and 559 pc. Hence,
stars with low $\sigma_z$ indeed belong to the disk, whereas values of $\sigma_z$ as high as about
100 Km/s, that implies in $z_0\approx1.4$ Kpc, are certainly not related to the disk component.
It is interesting to note that these results are in agreement with the quasi constancy of $z_0$
along galactic disks \citep{deg97,vdk02}. But note also that the radial rise in $z_0$ is more
expressive in the disks of early--type spirals.

Thus, as we have just shown, by measuring $\sigma_z$ we can have a direct estimate of the vertical extent
of the disk. A recently formed bar, as being part of the disk, still has the kinematical properties that
characterize disks, namely a low velocity dispersion, that produces a vertically thin structure, that
we can infer by observing low ($\lesssim$ 50 Km/s) $\sigma_z$ values. As the bar evolves, the already
mentioned processes like vertical resonances and the hose instability give the bar a higher
vertical extent by raising $\sigma_z$. These processes, however, make no changes in the remaining
of the disk.

In this way, we are able now to devise prescriptions that will make us able to distinguish between
recently formed and evolved bars. The first one is possible if one has measurements of $\sigma_z$ in the
bar and in the disk outside the bar and the bulge. This was possible for a few cases in this study, where
the outer spectra taken along the minor axis of the bar reach areas in the disk where the light
from the bar and the bulge makes only a small contribution. By comparing the velocity dispersion in the bar
and in the disk, a recently formed bar will have yet disk kinematics, whereas an evolved bar will have
a larger $\sigma_z$ than the disk. In Fig. 5 we show schematically how this comparison may
be done. The upper panel is an example of a recently formed bar, while the lower one shows the signature
of an evolved bar.

Since obtaining kinematical data for the fainter parts of the disk of galaxies is very telescope time demanding,
in most of the cases the outer spectra taken along the bar axes are still within the bar
or the bulge, even considering
the minor axis. Thus, we often may not have an estimate of $\sigma_z$ in the pure disk to compare to the
one in the bar. Hence, we expect that a young bar has not only a low $\sigma_z$ but also that its
radial profile shows a steep outward fall, as the bulge in the center has a much hotter kinematics.
Figure 6(a) shows schematically what would be the expected
signature of such a case. On the other hand, an old bar
is dynamically hotter than the disk, and is possibly as hot as the bulge, and in such case the
$\sigma_z$ radial profile is somewhat flat [see Fig. 6(d)]. However, we may expect to find cases
in which the bulge contribution may be misleading. Especially for late--type galaxies, whose bulges
may be substantially dynamically colder than average, even a young bar may display a fairly flat
$\sigma_z$ profile, as we show in Fig. 6(b). Furthermore, for the early--type galaxies with
their very hot bulges, an evolved bar may show up a steep $\sigma_z$ profile as can be seen
in Fig. 6(c).

Having set up the prescriptions to distinguish young and old bars we postpone to \S\S\,5 and 6 the
step further of evaluating how much older than young are evolved bars. Instead, in the next section
we present the results obtained with the spectra taken in this study.

\section{Results}

The  $\sigma_z$ radial profiles along the bar major and minor axes of the galaxies in our sample are
shown in Fig(s). 7 and 8 for the South and North samples, respectively.
The reason for a few empty panels in these figures is that we unfortunately were not
able to make measurements along the minor axis of the bars in NGC 4984 and NGC 5383.
Although we do not present and discuss our results concerning $h_3$ and $h_4$,
since the errors on these number estimates render them too unreliable,
the fact that similar results are obtained from both parameterizations (i.e., pure gaussian and
Gauss--Hermite series) is encouraging.
In order to properly analyse these profiles, and try to evaluate which bars are young and which are not,
it is also
important to know over which radial range the bulge dominates the emitted light and where the light
is dominated by the bar or the disk, whose kinematics we want to measure. To determine this we used the
{\sc aladin} interactive sky atlas \citep{bon00} with optical images publicly available through NED.
As estimates like these are somewhat subjective we did not try to establish them too precisely.
Nonetheless, these estimates serve indeed to our goals.
Also, the values referring to the bulge, in particular, mean where its light dominates over that from the bar
(and hence the measured kinematics) rather than been its length properly\footnote{Here we have assumed round bulges
for the sake of simplicity. The results from a detailed morphological analysis in \citet{gad05} for 7 galaxies
in our present sample indicate an average for the central bulge ellipticity $\approx$ 0.1. Hence this assumption
seems to be fairly reasonable to our purposes.}.
The results are shown in Table 2.
Note that our measurements along the bar major axis are always within the bar itself and that
the bulge dominates typically the inner 10 to 15 arcsec along both axes. We stress also that
since the spectra are extracted from ever increasing radial bins (from the center outward)
there is some overlap in radius that makes the transitions between bulge, bar, and sometimes
disk, smoother than in reality.

Following the discussion done so far and the results in Fig(s). 7 and 8 it is possible to qualitatively
assess bar ages and distinguish between young and old ones. A quantitative approach, however, is highly
desirable. Ideally, for this task one would like to have measurements of the
stellar vertical velocity dispersion
in the disk and in the bar free from {\em any} light contamination from other structural components,
especially the bulge. In this case, a simple comparison between the values of $\sigma_z$ would reveal
how much the bar has vertically evolved from the original disk.
Although this is feasible with the slit configuration we used it is virtually impossible
with 2-meter class telescopes, since this means obtaining high S/N spectra from very faint parts
of galaxies. Nevertheless, with the data presented here we can pursue such endeavor if we carefully
take into account the bulge contribution to the recorded light, using our estimates in Table 2. In addition,
let us define a fiducial value that represents $\sigma_z$ in the bar as
the average of our farthest measurements along the bar major axis, that for the majority of our galaxies
correspond to $\approx50\%-60\%$ of the bar semi-major length, $\sigma_{z,{\rm bar}}$ [column (6) in Table 2].
This choice comes from a compromise between minimizing bulge contamination and considering the galactocentric
distances our spectra reach. Similarly, we can define a fiducial $\sigma_z$ value for the disk.
Since galactic disks have generally a gradient in velocity dispersion we choose to take the disk fiducial
$\sigma_z$ at the same galactocentric distance from which $\sigma_{z,{\rm bar}}$ is defined. Only in
this way a meaningful comparison of both values can be achieved. Obviously, however, the disk velocity
dispersion is taken along the bar {\em minor} axis, and the galactocentric distance chosen corresponds
to about 1 to 2 times the bar semi-{\em minor} length. Again, we took an average of our measurements
at both sides of the center. Column 7 in Table 2 shows the difference between this and $\sigma_{z,{\rm bar}}$,
defined as $\Delta\sigma_z$. Hence, with the help of $\sigma_{z,{\rm bar}}$ and $\Delta\sigma_z$ we can make a more
quantitative assessment in order to distinguish young and old bars.

Then, according to our line of reasoning, some clear cases can be identified.
One can say with a certain degree of confidence that NGC 1326, NGC 4394 and NGC 5383
harbor recently formed bars. Their results resemble what we would expect based on Fig(s). 5(a) and 6(a)
(for NGC 1326 and NGC 5383) and 6(b) (for NGC 4394).
Moreover, $\sigma_{z,{\rm bar}} < 40$ Km/s and $\Delta\sigma_z < 10$ Km/s (although we do not have
this parameter for NGC 5383), that in fact point to these bars as being still a vertically thin structure,
belonging to the plane of the disk.

Clear instances of evolved bars are the ones in NGC 1302, NGC 1317 and NGC 5850. Their $\sigma_z$
radial profiles resemble Fig(s). 5(b) and 6(d). These galaxies have $\sigma_{z,{\rm bar}}$ equal
to 100, 145 and 60 Km/s. Although the latter value is clearly much lower, this is a result
from the fact that NGC 5850 is of a later morphological class than the former galaxies. In fact,
$\Delta\sigma_z \geq 30$ Km/s in the three cases, an evidence that these bars have vertically grown up
from their parent disks.

With these clear instances we can identify typical values for $\sigma_{z,{\rm bar}}$ and $\Delta\sigma_z$
in old and young bars. For the recently formed ones the average of $\sigma_{z,{\rm bar}} \approx 30$
Km/s where this climbs to $\approx$ 100 Km/s for the evolved ones. Similarly, $\Delta\sigma_z
\approx 5$ and 40 Km/s for young and old bars, respectively.

Let us go now to the less clear cases.
The behavior of $\sigma_z$ in NGC 1387 and NGC 1440 shows a relatively steep fall from the center
outward in both axes, and their $\Delta\sigma_z$ values are low, presumably indicating young bars.
However, their values for $\sigma_{z,{\rm bar}}$
are too high to be the case of young bars. These are certainly
vertically thick and evolved bars, resembling the schema of Fig. 6(c). These are typical cases where
the early--type bulge presents a very high velocity dispersion.
The low values of $\Delta\sigma_z$ may likely be explained by light contamination in the bar minor
axis from the bulge and the bar even at the outermost spectra. In fact, in these cases,
we have no reliable data on the velocity dispersion of the disk alone.
Moreover, note that in the outer
spectra $\sigma_z$ is as high as $\sim$ 150 Km/s. It is hard to devise how the global dynamical disk
instability alone could be responsible for the bar in these hot
disks, even taking into account that a substantial part of this high velocity dispersion
comes from the bulge and the bar.
NGC 4608 and NGC 5701 are similar cases. The bulge and bar influence may be causing low values for
$\Delta\sigma_z$, but the high values of $\sigma_{z,{\rm bar}}$ clearly put these bars in the
evolved bin. Their bulges are, however, not as dynamically hot as those of NGC 1387 and NGC 1440
and their $\sigma_z$ radial profiles resemble Fig. 6(d).

The following cases are more doubtful and thus the reader should have in mind that our conclusions
on these cases should be taken with care. NGC 4314 has intermediate values for both $\sigma_{z,{\rm bar}}$
and $\Delta\sigma_z$ and we do not reach far out enough along the bar major axis. However, the $\sigma_z$
radial profiles along the bar minor axis show an estimate of $\sigma_z$ in the disk substantially lower
than those that refer to the bar. Taking this single point into account it seems that the more
appropriate is to consider its bar an evolved one. It can in fact be an example where bar evolution
is at an intermediate stage. 
NGC 4579 has a $\sigma_{z,{\rm bar}}$ value closer to that of an evolved bar and a $\Delta\sigma_z$ value
typical of a young one. Our farthest measurements, however, clearly indicate that bar and disk have
a similar $\sigma_z$, at least in their outer parts, which makes us consider this bar as still
recently formed. Contamination from bulge light may again be the cause for the high $\sigma_{z,{\rm bar}}$
values. NGC 2665 is a similar case, and the very low and uncertain $\sigma_z$ estimates in the farthest
points along the bar minor axis may be occulting a young bar behind a high value for $\Delta\sigma_z$.
In fact, the $\sigma_z$ estimates along the bar minor axis do not agree in the outermost points from
each side of the galaxy center. Good spectra taken at farther galactocentric distances certainly
avoid and clarify these doubtful cases.

Finally, the case of NGC 4984 is the only one in that we could not arrive to a definite conclusion.
The dispersion reach relatively low values but only in one side of the bar. The lack of minor axis spectra
also prevents us to reach a conclusion. Note also that the inner structure of this galaxy is complex
(see Fig. 1), and that \citet{jun97} suggest that this galaxy has a secondary bar possibly almost
aligned with the primary. This complexity certainly is reflected in our results.

We performed three different statistical tests within the {\sc r} environment
(see http://www.\linebreak R-project.org)
to check if in fact the values we obtained for $\sigma_{z,{\rm bar}}$ and $\Delta\sigma_z$ to young
and old bars point to different objects. An unpaired Student t-test gives a 98\% probability that
what we defined by young and old bars are indeed different populations considering $\sigma_{z,{\rm bar}}$.
Support to this conclusion comes from Kolmogorov-Smirnov and Wilcoxon (or Mann-Whitney)
tests. These tests, however, do not find a significant difference between young and old bars if
one considers $\Delta\sigma_z$. As discussed above, the lack of $\sigma_z$ measurements far out
in the disk (away from the bar and bulge light) for several cases that an old bar is clearly seen
from the analysis of $\sigma_{z,{\rm bar}}$ is the likely cause of the latter result.
On the other hand, it is evident that the results from these tests considering $\sigma_{z,{\rm bar}}$
agree with the distinction made here between recently formed and evolved bars. In a later stage it would
also be interesting to try to identify intermediate cases. Note that tests like these do not take into account
the observational error in each single measurement of, e.g., $\sigma_{z,{\rm bar}}$. They, however,
use the standard deviation from the mean in both samples (in this case, young and evolved bars) to
estimate the observational error and then state the statistical significance of the final result,
i.e., the difference between the two samples.

In Fig(s). 7 and 8 one can also verify that at least in NGC 4314 and NGC 5850 there is a central drop
in the velocity dispersion. Doubtful cases may be NGC 1317, 4394, 4608, and NGC 5701, since
the drop, or the constancy, of $\sigma_z$ is marginally significant. The
remaining majority of the cases show a peak instead, that is more commonly observed. This peculiarity
has been also noted by \citet{ems01} in 3 other cases. This drop may be caused by a recently
formed stellar inner disk originated from the funneling of gas to the center by the bar. In this
context, it is noticeable that NGC 4314, NGC 4394, and NGC 5850 present indeed inner disks,
as we confirm it through a detailed structural analysis in \citet[hereafter Paper II]{gad05} using the
{\sc budda} code \citep{des04}.

Many examples of measurements of the stellar velocity dispersion in galaxies can be found in the
literature but we are not aware of a systematic measure of this physical parameter along face--on
bars. Nonetheless, comparisons with previous values may be instructive, even if not exactly relative
to a similar study. \citet{cor03} made measurements of the stellar velocity dispersion in NGC 4984
in a manner similar to what we present here. A comparison of the results from both studies shows
that the estimates are essentially the same.
\citet{fis97} shows estimates for the stellar kinematics of lenticular galaxies.
One can verify that his measurements are very similar to ours as his velocity dispersion estimates
for, e.g., NGC 3412, 3941 and NGC 4754 (all barred and more or less face--on) range from
$\approx$ 200 Km/s in the center to $\approx$ 60 Km/s 20 arcsec away from it. NGC 3941 is
an interesting case worthy to be explored in more detail. Not considering that his measurements are
at an angle to the bar major axis one could use our analysis to conclude that the bar is evolved,
although better estimates for the velocity dispersion in the disk would be necessary to a more
clear conclusion. A similar study involving late--type galaxies can be found in \citet{piz04}.
Looking their results for galaxies which are reasonably face--on we again find measurements
similar to ours. This is the case for NGC 210, 3054, 6878 and NGC 7412, whose stellar velocity
dispersion ranges from $\approx$ 150 Km/s in the center to $\approx$ 50 Km/s 25 arcsec away from it,
although we would expect somewhat lower values in the disks of the two latter galaxies.
Similar results were found for NGC 488 and NGC 2985 \citep{ger97,ger00} and in \citet{bot93}
one also finds, for the four face--on galaxies in his sample, estimates similar to ours.
If one considers that $\sigma_z/\sigma_\varphi \sim 0.83$ then inspecting
the recent measurements of \citet{kre04} one sees that the range in velocity dispersion they observe
in the disks in a sample of edge--on late--type galaxies is in agreement with our estimates for both
the late--type galaxies in our sample and the early--type ones, whose $\sigma$ reaches considerably
higher values, also in agreement with previous work (see \S\,3.2 above).

Table 2 also shows error estimates for $\Delta\sigma_z$. We followed two different methods to determine
these errors. In the first one we propagate the error as usually done in error analysis considering the
errors from the spectrum fitting, i.e., those errors determined in \S\,3.1, displayed as error bars in
Fig(s). 7 and 8. In our second method we do not consider the uncertainties in the spectrum fitting, but
assume that the error in $\Delta\sigma_z$ is the quadratic mean of the differences between the values
of $\sigma_z$ in the bar and in the disk (that are used to calculated $\Delta\sigma_z$) taken at
both of sides of the galaxy center. This sometimes leads to smaller errors. This analysis shows
that the relative error in $\Delta\sigma_z$ is large, as one should expect considering it is a 
difference between close numbers, each one already bringing along considerable uncertainty, since
they are estimated from faint light. The impact of this uncertainty in our division between
young and evolved bars must not be exaggerated, however, since this was done taken also into account
$\sigma_{z,{\rm bar}}$ and the full $\sigma_z$ radial profiles.

\section{The Vertical Thickening of Bars in $N$--Body Experiments}

It is interesting now to compare the results described in the last section with $N$--body realizations
of the evolution of barred galaxies. This is useful, for instance, to establish the time scales involved
in the vertical thickening of bars. Previous studies estimate the time scale for the occurrence
of the boxy--peanut morphology, likely a consequence of vertical resonances and/or the
hose instability, in the order of 1 Gyr \citep[see, e.g.,][]{com81,com90}.
But a systematic comparison of the vertical velocity dispersion in
observations and simulations has not been done yet. In this section we perform simplistic
$N$--body experiments on the evolution of pure stellar disks
to use as a first--order approach in making this evaluation.
A more realistic treatment would involve adding responsive bulge and halo components, which
is beyond our present scope. We note, however, that we intend to perform a more thorough
and accurate analysis in a future paper.
Thus the results from this section must be considered with caution and tested with
more realistic experiments. They are, however, useful to obtain an approximate
order of magnitude of how much is the age difference between the young and evolved bars
identified above, as well as, and this may be even more important, to attest that measurements
of the vertical velocity dispersion along bar and disk are, at least qualitatively, useful
for bar age estimates, as we propose here.

The experiments were done within the {\sc nemo} package \citep{teu95}, and the \citet{bar86}
tree--code was used for force calculation. We used virial units, in which $G=M=-4E=1$, where $G$
is the gravitational constant, $M$ is the mass and $E$ is the total energy of the system. In the runs
a constant time step of typically $2\times10^5$ years was employed, giving about 10$^3$ time steps
per crossing time in the systems analysed. The optimal softening parameter suggested by
\citet[see also \citealt{ath00}]{mer96} for $N=10^5$,
that is the number of particles we have used typically,
is in the range from $\epsilon\approx0.01-0.05$, and the latter value was used.
This has the intention of minimizing spurious effects caused by the low number of particles, especially
two--body relaxation that may heat and thicken the disk.
Finally, the aperture (or tolerance) angle used was always $\theta=0.7$.

The simulated disks are made of responsive particles of equal masses and
are exponential and isothermal with a constant height scale \citep{fre70,fre78,vdk81}:

\begin{equation}
\rho_d(r,z)=\frac{M_d}{4\pi R_d^2 z_0}e^{-r/R_d}{\rm sech}^2(z/z_0),
\end{equation}

\noindent where $M_d$ is the disk mass, being $R_d$ its length scale. In this case, the Toomre $Q$
parameter, establishing if the disk is unstable to the bar mode instability, is defined as it is done
usually:

\begin{equation}
Q\equiv\frac{\sigma_r\kappa}{3.36G\Sigma},
\end{equation}

\noindent where $\kappa$ is the epicycle frequency of the stellar orbits in the plane of the disk, and
$\Sigma$ is the projected mass surface density. Thus, since
$Q\propto\sigma_r=2\sigma_z\propto\sqrt{\pi\Sigma z_0}$ \citep[see, e.g.,][]{vdk02},
the stability of the disk is set up in the initial conditions for $z_0$. In some of the experiments,
however, we have directly attributed a constant value to $Q$ in the initial conditions, aiming
to force the bar instability. In the latter case, is the vertical extent of the disk that is determined
as a function of the Toomre parameter. A correction for the asymmetric drift was applied in the
disks, assuming the Milky Way disk as typical \citep[see][]{deh98}.

Several simulations were run as to mimic the 2 Gyr evolution of stellar disks with properties
similar to the Milky Way disk \citep[see][]{bin87,bin98}, i.e., a mass $M_d=6\times10^{10}$ M$_\odot$
and length scale $R_d=3.5$ Kpc. The height scale varies in the different experiments
in the range $z_0=200-600$ pc, which, as discussed above, have fundamental influences on the
disk stability against bar formation.
The center of mass and energy variation were typically of the order of 0.3\%.

We now discuss, based on one representative experiment
(with $z_0=450$ pc), the onset of the bar instability in galaxies
and the vertical thickening of bars.
This fiducial calculation has a Toomre parameter in the initial conditions estimated by Eq. (15) that decreases
continuously from $\approx$ 3 in the center to $\approx$ 1 at the outermost radii.
Figure 9 displays the evolution of the pure stellar disk in that, as
expected, a strong bar develops. This happens after $t=6\times10^8$ yr and spiral arms also appear
in the so called grand design morphology. The bar weakens after $t=5\times10^8$ yr but remains
for a similar period, then giving place to an oval bulge--like
distortion until the end of the simulation after 2 Gyr. It can be
seen clearly in the edge--on projection the vertical heating of the disk and how the weakening of the bar
originates a bulge--like structure, corroborating previous works \citep[see also][]{deb04}.
It is, however, still a matter of debate whether bars are robust, perennial galactic components,
or may be easy to dissolve with central mass concentrations, as a result of their natural
evolution \citep[see][]{she04}.

Figure 10 shows the evolution of the rotation curve in our experiment. This was derived
through the simulation of long slit spectroscopy data, using parameters that match
the observations we present in \S\,2, in what concerns the slit width and pixel size on the ``sky''
and the seeing, i.e., spectral and spatial resolution. The same procedures were done to extract
radial profiles of the stellar vertical velocity dispersion along
the major and minor axes of the bar, which will
be presented and discussed shortly. Note that the rotation velocities reached are low due to the
fact that we are not including the mass contributions of either bulge or halo. However, the global shape
of the rotation curve is similar to what is generally found for real galaxies \citep[e.g.,][]{rub85}.
Interestingly, although there is no dark matter halo contribution to these curves, they are
quite flat: only a hint of a fall is observed in the second row of panels. This is at least partially
explained by the fact that the particle distribution in our disk models is truncated at about
10 Kpc, i.e., only $\approx$ 3 length scales \citep[but see][]{bar04}.

We have also made some tests with live bulges and rigid halos as to realize more
complete and realistic systems, and verified the difficulties encountered by the old
bar formation scenario (that based purely on the bar mode disk instability) in
explaining the origin of bars in galaxies with kinematically hot disks and conspicuous
bulges, as observed barred lenticulars.
However, we can identify five possible ways out that need to be fully exploited
to salvage the bar mode instability in disks as the origin of bars in
galaxies. First, as showed recently by \citet{ath02a} and
\citet{ath02b,ath03}, the use of rigid halos is misleading,
as their simulations with live halos show that the exchange of angular momentum between the disk
and the halo may, contrary to expectations led by previous studies, reinforce the bar.
Hence, the present standard bar formation scenario adds to the bar mode disk instability
the absorption of angular momentum by a responsive halo. Second,
as in the scenario of \citet{bou02}, the infall of gas from the halo in the disk may also create
a kinematically cold structure that is bar unstable. Third, as we showed in \citet{gad03}, sufficiently
eccentric halos can induce bar formation in kinematically hot stellar systems. Finally, we shall also
note that tidal interactions may play a relevant role in this context \citep{nog00} and the
approach of \citet{pol03}.

In all our experiments that do not preclude the appearance of a bar, its properties match previous
studies, both in theory and observation. For example, all bars develop quickly in a few times
$10^8$ yr and have a length in the range from $\approx$ 4 to $\approx$ 8 Kpc. We have also
estimated the bar pattern rotation velocity to be around 30 Km s$^{-1}$ Kpc$^{-1}$, in general.

Concerning the vertical extent of bars, we show in Fig. 11 that, as predicted, bars grow vertically
thick as they evolve, producing a structure similar to a bulge seen edge--on, also presenting the characteristic
boxy--peanut morphology. At the time this structure is present, the bar no longer belongs to the disk, since
the growth in $\sigma_z$ produces the vertically thick pseudo bulge. Furthermore, our simulations
indicate that this morphology develops quickly (in a few dynamical times, i.e., a few times $10^8$ yr after the
formation of the bar) and weakens after $\sim$ 1 Gyr. Thus, the boxy--peanut morphology may be used
as an indication of a young bar, i.e., one that has been originated in less than about 1 Gyr ago. Note also
that the presence of a pre--existent bulge would turn difficult the proper identification
of the boxy--peanut morphology.

However, in our simulations, $\sigma_z$ in the bar does not exceed 50 Km/s during the boxy--peanut
phase and not even after 2 Gyr of bar evolution. This is shown in Fig. 12, where we present the evolution
in time of the stellar vertical velocity dispersion along the bar major and minor axes
of our numerical experiment.
These profiles were derived using the same procedures we used to produce the rotation curves
discussed above. Although tentative,
this is evidently in contrast to the observations we reported in the previous section,
where evolved bars have $\sigma_z\sim$ 100 Km/s. Thus, while
the processes that create the boxy--peanut morphology are fast, the one(s) responsible for the
{\em observed} vertically thickening of the bars should happen in a longer timescale.

A qualitative analysis of Fig. 12 proves very useful. In the first panel (at $t=0$), corresponding to the
initial conditions, the kinematics along major and minor axis of the disk should be
(nearly) identical (with small differences from statistical fluctuations), as evolution
yet did not start. To produce the velocity dispersion radial profile along the minor axis, we introduce
an error of 5 degrees in the slit position angle and an error within the seeing in its
central position. This allows one to note that indeed the kinematics is similar in both axes, and
also to evaluate how these uncertainties, which are often present in real observations, would affect
the measurements. At $t=200$ the bar is recently formed and, as we argued above, keeps
its kinematical properties along the vertical axis similar to those of the disk. At $t=400$ and
$t=600$, however, the signature of the evolved bar is clear: in the region dominated by the bar
(whose semi-major axis length is about 4 Kpc)
$\sigma_z$ is substantially higher than in the bar--free region of the disk. Hence, our simulations give
at least a qualitative support to the bar age diagnostic tool we introduced above. If bars are robust
they may keep this signature and yet maybe acquire higher values for $\sigma_z$, as we discuss
in the next section. In our simulation, however, the bar weakened, a process that starts at $t=800$
and goes until $t=1000$ giving origin to an unbarred galaxy with
a central morphologically bulge--like structure that also resembles
bulges form the kinematical point of view.

For the typical distance of the galaxies in our sample (cz $\sim1500$ Km/s) 10 arcsec corresponds
roughly to 1 Kpc, which means that our spectroscopic measurements presented in \S\,4 go typically
as far as about 2 Kpc from the center, where our simulations (Fig. 12)
show that a substantial difference in $\sigma_z$ would be observed between the
evolved bar and the disk.

\section{Discussion}

The ages of bars diagnostic we present here allows us to distinguish between young and
old bars, but a further question is to know how much is that age difference. The results from the
previous sections suggest that it is not only the processes that cause the boxy--peanut morphology that
produce bars as vertically thick as observed. This means that the age difference between young and old
bars may be significantly in excess of 1 Gyr. We suggest now that another mechanism is taking place
during the evolution of the bar, and that it goes on after the vertical resonances and the hose instability end,
giving the bar a vertical extent and a $\sigma_z$ in agreement with what our observations show.

As shown by \citet{spi51,spi53}, collisions with giant molecular clouds (GMCs) are able to make the disk
gradually hotter and vertically thicker. This is a process that happens in the disk as a whole
\citep[e.g.,][]{bin01}, but we can reasonably assume that these clouds may be more concentrated along
the bar than in the remaining of the disk, given that bars are indeed able to collect and funnel the
gas present in the disk to the center. If this is true, then the effect of these collisions will be stronger
in the bar region of the disk. It should be mentioned also that this effect happens too with any other
inhomogeneity in the stellar density distribution.

The variation in the stellar velocities provoked by the impact with GMCs may be written in the impulsive
approximation as:

\begin{equation}
(\Delta v)^2=\left(\frac{2GM}{bv}\right)^2,
\end{equation}

\noindent where $M$ is the typical GMC mass, and $b$ is the impact parameter. The succession of
several encounters gives origin to a process of diffusion in phase space \citep{wie77} in the form:

\begin{equation}
{\rm d}\sigma_z^2=D(\Delta v_z){\rm d}t,
\end{equation}

\noindent where $D(\Delta v_z)$ is the diffusion coefficient \citep[see][]{bin87}, proportional to
$\left(\frac{M}{bv}\right)^2$. If $D(\Delta v_z)$ is constant, then:

\begin{equation}
\sigma_z^2=\sigma_{0z}^2+D(\Delta v_z)t,
\end{equation}

\noindent where $\sigma_{0z}$ is the initial vertical velocity dispersion. Through numerical simulations,
\citet{vil83} showed that the equation above may be parameterized in the form:

\begin{equation}
\sigma_z=\sigma_{0z}\left(1+\frac{t}{\tau}\right)^n,
\end{equation}

\noindent with $n=1/2$ for a random walk diffusion, and where $\tau$ is a time scale proportional
to $b^2/M^2$. This value for $n$ is in agreement with the relation between the velocity dispersion
and the age of stars in the solar neighborhood.

Following the results from \citet{vil83}, this process may elevate the velocity dispersion from
$\sigma_z\simeq5$ Km/s to $\sigma_z\simeq25$ Km/s in about 7 Gyr. Assuming that GMCs
are accumulated along the bar, and the fact that $\tau\propto b^2$, this mechanism may explain
the observed vertically thickening of bars in timescales larger than 1 Gyr. If $b$ along the bar is half
the value of $b$ in the remaining of the disk, resulting from a higher concentration of GMCs in the
bar, then in 7 Gyr $\sigma_z$ in the bar might go from $\sigma_z\simeq5$ Km/s to $\sigma_z\simeq100$ Km/s,
as observed.

If we consider that during the vertical thickening process bars cease to be kinematically cold then we will
have to take into account that the diffusion coefficient is not constant and that ${\rm d}\sigma^2/{\rm d}t
\propto 1/\sigma$. This means that the scattering of stars by GMCs get less efficient in time when we have
two kinematically distinct components: the cold disk where GMCs fall, and the vertically rising and hotter
bar. This is certainly more justified from the theoretical point of view and leads to $n=1/3$,
although the resulting differences are small. Following again the results from
\citet{vil83} this means that the bar takes more
$1-2$ Gyr to achieve $\sigma_z\simeq100$ Km/s.

The results from the simulations presented in \citet{ath92a,ath92b} and \citet{pat00} may, however,
represent a caveat to the analysis above. This is because they show that gas concentrate in
relatively narrow strips along the leading edges of bars. If GMCs behave like gas does in their
simulations then the consequences on the stellar dynamics within the bar may be different
from what it would be if, as we assumed above, GMCs are uniformly distributed within the bar.
One should note
however that, although dust lanes tend in fact to delineate strips along the leading edges of bars
in real galaxies, there are also many examples of dust patches {\em across} bars
\citep[see, e.g.,][]{sel93}.

It is worth noticing at this point that in Paper II we have measured the optical colors of bars in a
sample of galaxies similar to the one in this work, including 7 galaxies\footnote{These are NGC 4314,
4394, 4579, 4608, 5383, 5701 and NGC 5850 -- i.e., the North sample -- which give us 3 recently formed
bars and 4 evolved ones.} studied here. We found
that the average value for B--V in the bars that our diagnostic identify as evolved is 1.1.
In contrast, the identified young bars have B--V equal to 0.7. This color difference may be translated
to an age difference of the order of 10 Gyr \citep[see, e.g.,][]{tin76,mar98}, even allowing for reasonable
differences in the metallicity of the stellar population. Although the age of the stellar population
in the bar does not necessarily represent the age of the bar, this result, in agreement of what we found
above with the numerical simulations, corroborate our suggestion that the age difference between
young and old bars is significantly higher than 1 Gyr. Although we are here neglecting reddening by dust
this is justified by the fact that we defined the bar color at its ends where dust effects
are minimized. There is also no reason to believe that dust attenuation should be more effective in
galaxies with evolved bars rather than in those with recently formed bars. The reader is referred to Paper II
to a more detailed discussion on these color comparisons. Interestingly, in this paper we also find that
there are no significant differences in the photometric shape of young and old bars. Considering the
7 galaxies in common
with the photometric analysis in Paper II, all their bars show a flat surface brightness
profile along the major axis, typical of barred galaxies with morphological types earlier than
$\approx$ Sbc \citep{elm85}. The elliptically averaged surface brightness radial profiles of the galaxies
also do not show any clear trend amongst young or old bars concerning profiles of Freeman types I and II
\citep{fre70}.

Within this line of reasoning
the evolved bars we identified in \S\,4 have ages not much lower than the age of the
universe and hence have never dissolved nor are recurrent. Obviously, this does not exclude
the possibility of recurrent bars in other galaxies, but it also reinforces the results from \citet{she04},
who suggest that bar dissolution requires much more mass concentration than observed, and thus
that bars are robust.

Using our diagnostic tool we showed in \S\,4 that from the 14 galaxies in our sample, 8 have
evolved bars whereas 5 have recently formed bars. This result may also be interpreted as
an observational evidence of the vertical thickening of bars. Evidently, a more direct way
to evaluate the vertical structure of bars is by observing them in edge--on galaxies. In this
case, however, the proper identification of bars is still difficult and also involves careful and detailed
measurements. Furthermore, in edge--on galaxies one is not able to retrieve several fundamental
galactic properties and their detailed structural characteristics.

Our ability in estimate the ages of bars is an important step forward in the study of the formation and
evolution of galaxies. We can, for instance, determine more precisely what are the time scales in
the secular evolution processes related to bars, and if a given barred galaxy have already suffered
from them or not. In fact, we see that from the 8 galaxies with evolved bars only two (25\%) have AGN,
whereas from the 5 galaxies with young bars three (60\%) show AGN activity. While this is yet
a small number statistics, it indicates
that the time scale for the fueling of AGN by bars is short. In this case, one may find a better
correlation between the presence of bars and AGN in galaxies if considering only young bars.
This result is also corroborated by those from Paper II, that show that the average B--I color in the bars of
AGN galaxies is 1.7, while it is 2.1 in galaxies without nuclear activity. One word of caution, however:
the significance of these results are debatable since AGN classifications are derived by different authors
in different ways.

It is also interesting to note that from the 8 evolved bars, 7 (88\%) reside in early--type galaxies (S0--Sa)
while only 1 (12\%) is in a Sb galaxy. On the contrary, the young bars seem to preferentially inhabit
later--type galaxies: from the 5 young bars, 2 (40\%) are found in S0--Sa galaxies and 3 (60\%) in Sb.
This point is further explored in Paper II. It is also important to stress
again that these results must be confirmed
by studies with much larger samples as the small number statistics here means their statistical
significance is low.

\section{Conclusion}

By exploring the predicted vertical thickening of bars during their evolution, through kinematical
measurements in the vertical direction of bars in a sample of 14 galaxies, along their major and minor axes,
we have developed and introduced a diagnostic tool that allows one to distinguish between
recently formed and evolved bars. Using several $N$--body realizations of bar unstable disk galaxies,
we studied the time scale involved in the vertical evolution of bars, and the conditions necessary
to the onset of the bar instability.

We found that bars may be broadly classified between young and old bars, by, essentially, comparing
the vertical velocity dispersion in the bar and in the disk in that it formed.
Young bars are kinematically similar (in the vertical direction) to
the disks they reside in, having a low vertical velocity dispersion,
thus being still a vertically thin structure in the
disk. Old bars, however, have values for vertical velocity dispersion significantly higher than those
in their disks, and hence are a vertically thick structure that does not belong to the disk anymore.

We present evidences that the time scale for the vertical thickening of bars may be substantially
larger than 1 Gyr, and thus that another physical process in addition to the vertical resonances
and the hose instability may be also playing a major role in this context. These evidences come,
first, from the numerical experiments, that show that even after 2 Gyr of evolution the simulated
bars have a vertical velocity dispersion that does not exceed 50 Km/s, whereas in our observed
bars we have measured values of the order of 100 Km/s. Second, this result is corroborated by those
we present in Paper II, showing that the average color difference between vertically thin and
thick bars suggest an age difference of about 10 Gyr. These results reinforce our suggestion that
the Spitzer--Schwarzchild mechanism is responsible for the later vertical thickening of bars, since
we demonstrated that it can produce the observed values for the vertical velocity dispersion
in these longer time scales. Furthermore, our simulations also show that the boxy--peanut
morphology appears quickly, in a few dynamical times after the formation of the bar, and
also dissolves relatively quickly, after about 1 Gyr, being hence an indication of a recently formed bar.

To estimate the ages of bars is evidently an important goal in the study of the formation and evolution
of galaxies, as it has been proved many times now the major role bars play in this matter. In this paper,
we already showed and discussed how it can help us to better understand the intricate relationship
between bars and the fueling of AGN. In Paper II, we use this tool together with multiband imaging
data to explore furthermore the impact of the formation and evolution of bars on the formation and
evolution of galaxies. It is highly desirable now to obtain high quality kinematical data along bars
in a growing number of galaxies, and also try to extend them to the outer bar and disk
limits, especially to avoid the effects of a bulge.
While we are now able to separate bars that are recently formed from the ones that have been already
evolving for time scales of several Gyr, the next relevant step is to develop a theoretical understanding
of the processes involved in the vertical evolution of bars, in order to be able to refine these age
estimates.

\acknowledgments
This work was financially supported by FAPESP grants 99/07492-7
and 00/06695-0. We thank Rob Kennicutt for his kind hospitality during a visit to the Steward Observatory
when part of this work was undertaken.  It is a pleasure to thank Lia Athanassoula for her detailed reading and
comments on a first version of this manuscript, in particular on its theoretical aspects, which helped to
improve our work. Comments and suggestions from both the observational and theoretical anonymous
referees were invaluable, and helped to substantially improve the presentation of our results.
We would like to thank Peter Teuben for help on using {\sc nemo} (\url{http://bima.astro.umd.edu/nemo/}).
This research has made use of Aladin and of the
NASA/IPAC Extragalactic Database (NED), which is operated by the Jet Propulsion
Laboratory, California Institute of Technology, under contract with the National Aeronautics and Space
Administration. This research has also made use of NASA's Astrophysics Data System, the Lyon
Extragalactic Data Archive (LEDA, \url{http://leda.univ-lyon1.fr/}), and of spectral data retrieved from the
Elodie archive at the Observatoire de Haute-Provence (OHP, \url{http://atlas.obs-hp.fr/elodie}).
Statistical tests were made within the {\sc r} environment (http://www.R-project.org).

\clearpage

\begin{deluxetable}{llcccccc}
\tablecaption{Basic data for all galaxies in our sample.}
\tablewidth{0pt}
\tablehead{Name & Type & D$_{25}$ & log R$_{25}$ & m$_{\rm B}$ & cz & AGN & Companion \\
(1) & (2) & (3) & (4) & (5) & (6) & (7) & (8)}
\startdata
NGC 1302    & SB0(r)          & 3.89         & 0.02         & 11.40          &     1730      & \dots                & N     \\
NGC 1317    & SABa(r)        & 2.75          & 0.06         & 11.85           &    1941     & \dots                & Y     \\
NGC 1326    & SB0(r)          & 3.89          & 0.13         & 11.42          &    1365      & LINER              & N     \\
NGC 1387    & SAB0(s)       & 2.82          & 0.00         & 11.82          &    1328      & \dots                & Y     \\
NGC 1440    & SB0(rs)        & 2.14          & 0.12         & 12.58          &    1504      & \dots                & N     \\
NGC 2665    & SBa(rs)        & 2.04          & 0.13         & 12.96          &     1740      & \dots               & N     \\
NGC 4314    & SBa(rs)        & 4.17          & 0.05         & 11.22          &     963       & LINER              & N     \\
NGC 4394    & SBb(r)          & 3.63          & 0.05         & 11.53          &     772       & LINER              & Y     \\
NGC 4579    & SABb(rs)      & 5.89          & 0.10         & 10.68          &    1627      & LINER/Sey1.9  & N     \\
NGC 4608    & SB0(r)          & 3.24          & 0.08         & 11.96          &    1823      & \dots                & N     \\
NGC 4984    & SAB0(rs)      & 2.75          & 0.10         & 11.80          &    1243      & \dots                & N     \\
NGC 5383    & SBb(rs)        & 3.16          & 0.07         & 12.18          &    2226      & \dots                & Y      \\
NGC 5701    & SB0/a(rs)     & 4.26           & 0.02         & 11.82          &    1556      & LINER             & N      \\
NGC 5850    & SBb(r)          & 4.26           & 0.06         & 12.04          &    2483      & \dots               & N     \\
\enddata
\tablecomments{Columns (1) and (2) show, respectively, the name and the morphological type of the
galaxy, while column (3) shows its diameter in arcminutes at the 25 B magnitude isophotal level, and
column (4) shows the decimal logarithm of its major to minor axes ratio at the same level. Columns (5)
and (6) show, respectively, the apparent B magnitude and the radial velocity in Km/s. All these data were
taken from \citet[hereafter RC3]{dev91}. Column (7) presents an AGN classification according
to the NASA Extragalactic Database (hereafter NED). In column (8), ``Y'' means that there is a companion
galaxy similar in size physically interacting within 10 arcminutes, while ``N'' means that there are
no companion galaxies. To make this analysis we used the RC3 and the Lyon Extragalactic Data
Archive (hereafter LEDA).}
\end{deluxetable}

\clearpage

\begin{deluxetable}{lccccccccl}
\rotate
\tabletypesize{\scriptsize}
\tablecaption{Relevant properties of bulges and bars of the galaxies in our sample to a proper
evaluation of the bar ages from the radial profiles of the vertical velocity dispersion
[Fig(s). 7 and 8].}
\tablewidth{0pt}
\tablehead{Galaxy & Bulge & Bar & Bulge & Bar & $\sigma_{z,{\rm bar}}$ & $\Delta\sigma_z$ & Error (1) & Error (2) & Bar \\
NGC & Major Axis & Major Axis & Minor Axis & Minor Axis & (Km/s) & (Km/s) & (Km/s) & (Km/s) & Age \\
(1) & (2) & (3) & (4) & (5) & (6) & (7) & (8) & (9) & (10)}
\startdata
1302        & 10      &    40        &   10             &  20    & 100    & 28 & 47 & 12 & old                \\
1317        &  15     &    35        &   15             &  25    & 145    & 57 & 71 & 70 & old                \\
1326        &  10     &    35        &   10             &  10    & 38    & 8       & 34 & 6 & young                \\
1387        &  15     &    45        &   15             &  30    & 159    & 2  & 52 & 46 & old                \\
1440        &  15     &    35        &   15             &  15    & 178    & 2      & 83 & 20 & old                \\
2665        &   10    &    35        &   10             &  20    & 67    & 32  & 43 & 66 & young?                \\
4314        &  10     &    80        &   10             &  15    & 59$^a$    & 20  & 21 & 32 & old?                \\
4394        &  7       &    40        &   7               &  10 & 28    & 0        & 27 & 23 & young                \\
4579        & 10      &    40        &   10             &  10   & 57    & 6        & 20 & 3 & young                \\
4608        & 15      &    45        &   15             &  10   & 62    & 9        & 38 & 35 & old                \\
4984        &  15     &    35        &   \dots         &  \dots   & 68    & \dots  & \dots & \dots & \dots                \\
5383        &  10     &    52        &   \dots         &  \dots  & 23$^a$    & \dots   & \dots & \dots & young                \\
5701        &  15     &    50        &   15             &  10    & 85$^a$    & 5       & 52 & 7 & old                \\
5850        &  8       &    72        &   8               &  12  & 60$^a$    & 33  & 40 & 4 & old                \\
\enddata
\tablecomments{\tiny Column (1) gives the NGC number of the galaxy while columns (2) through (5) show
the apparent length in arcseconds of the semi-major and semi-minor axes of bulges and bars after
visual inspection (see text for details). Note that our measurements reach approximately 20 arcseconds
from the center. Column (6) gives the vertical velocity dispersion measured
along the bar major axis at $\approx
50\%-60\%$ of the bar semi-major length, unless when $^a$ notifies that our spectra do not reach this 
galactocentric distance. Column (7) gives the difference between columns (6) and
$\sigma_z$ measured in the disk along the bar minor axis at the same galactocentric distance.  
The values in columns (6) and (7) are averages from both sides of the bar axes.
Columns (8) and (9) show error estimates in the values of $\Delta\sigma_z$ according to two
different procedures (see \S\,4). Finally, column (10) shows whether the galaxy harbor an
young or and old bar according to the method and criteria developed here. Doubtful cases are notified.}
\end{deluxetable}

\clearpage

\begin{figure}
\includegraphics[width=4cm,height=4cm,keepaspectratio=false]{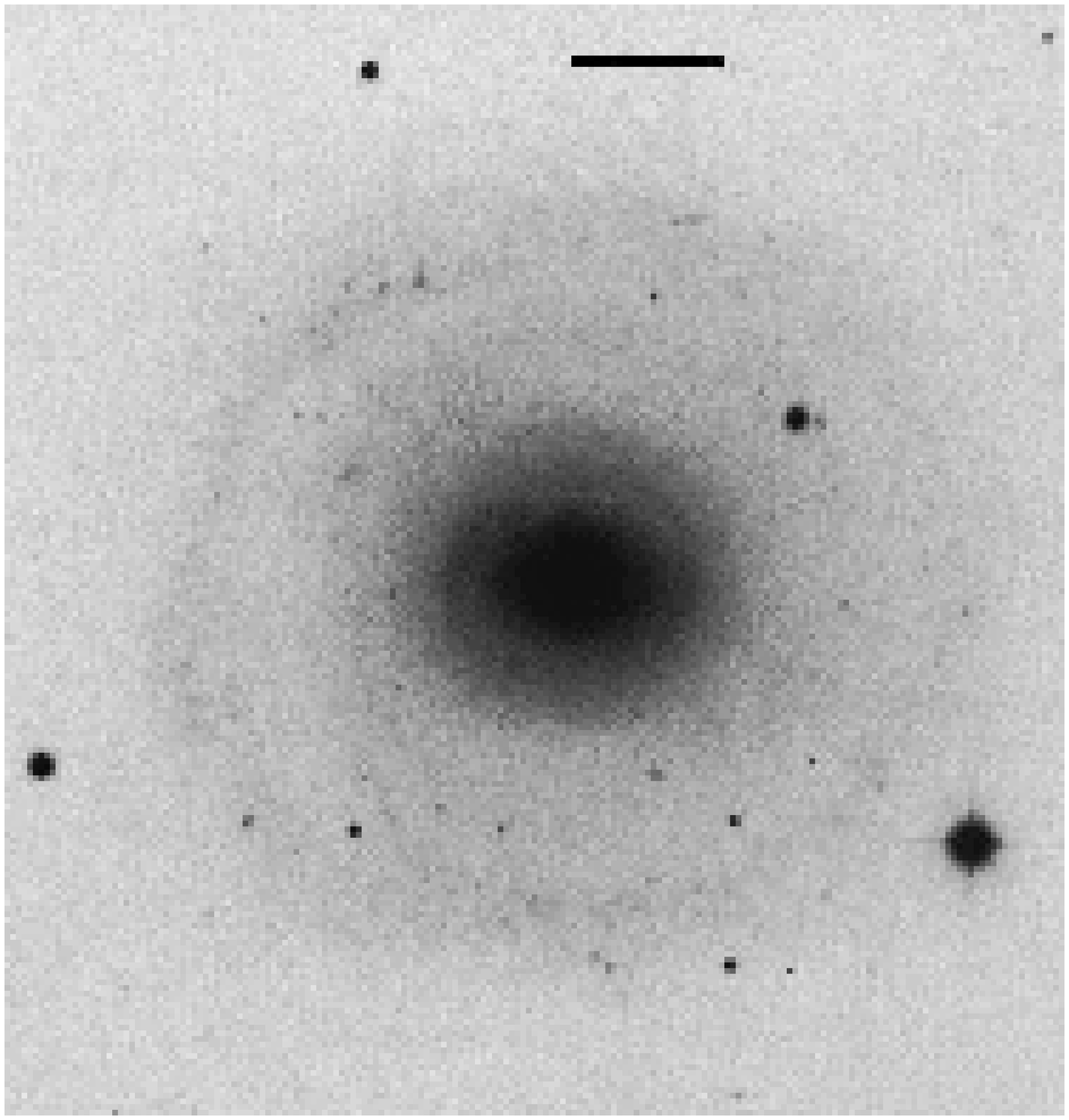}
\includegraphics[width=4cm,height=4cm,keepaspectratio=false]{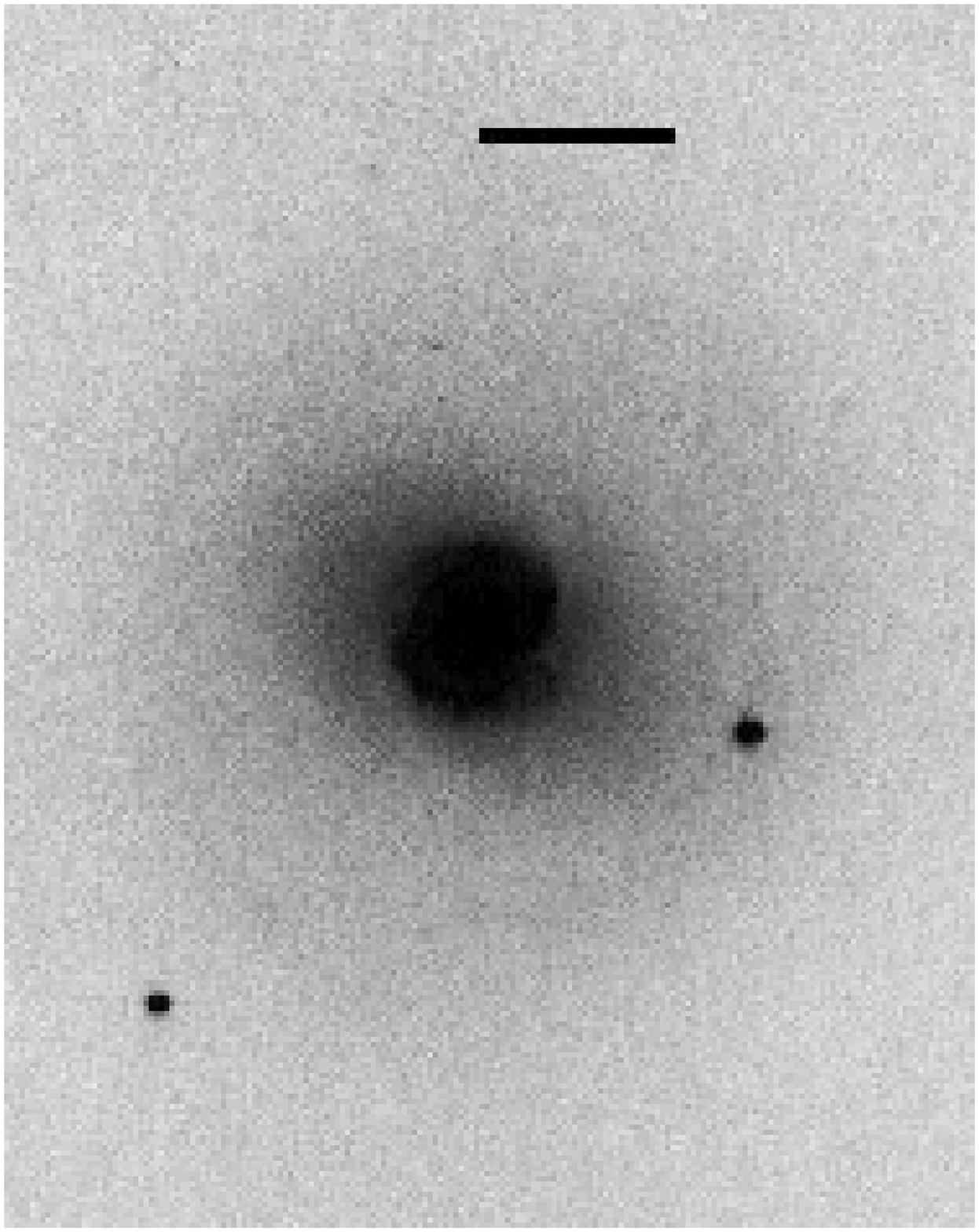}
\includegraphics[width=4cm,height=4cm,keepaspectratio=false]{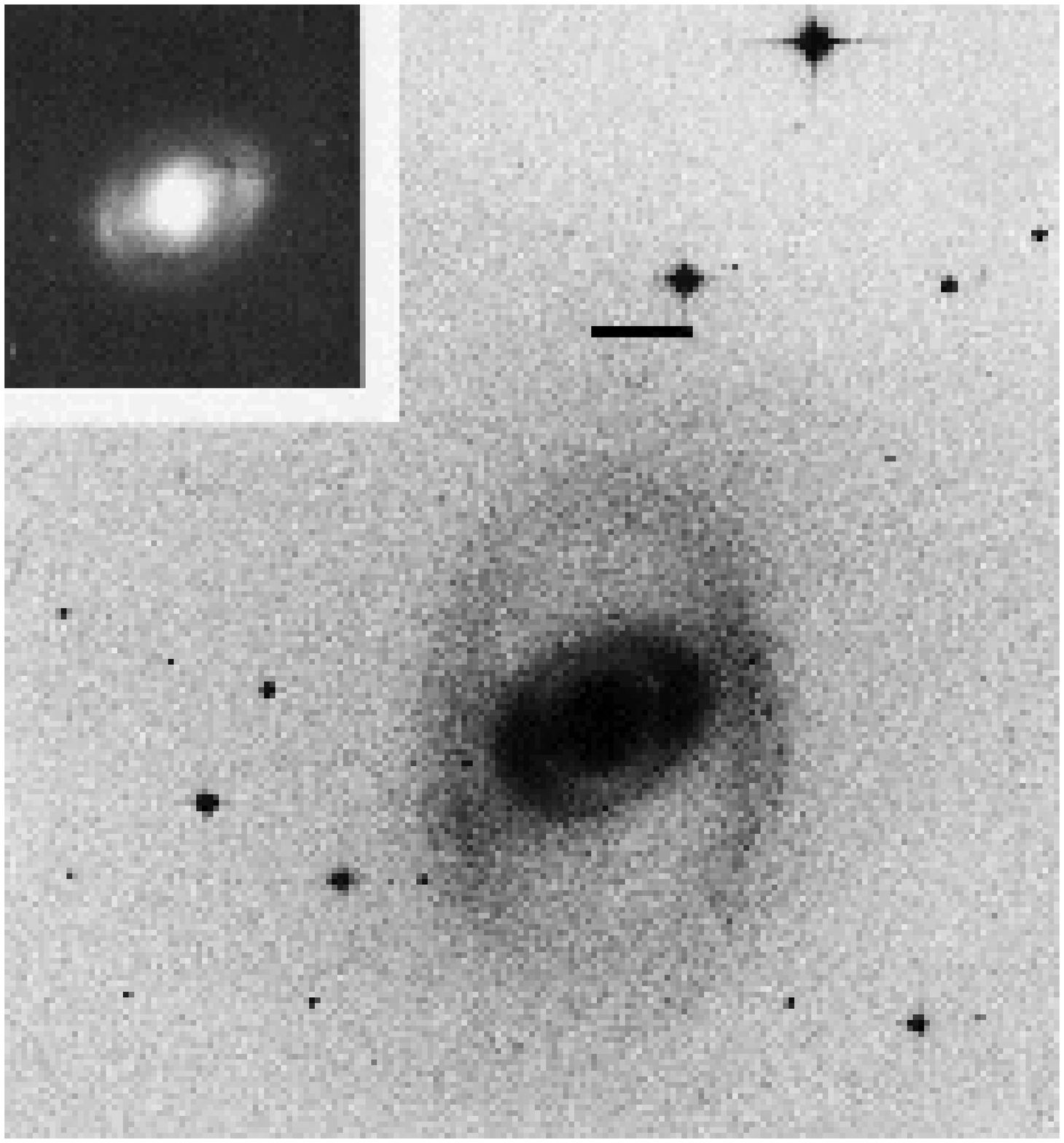}
\includegraphics[width=4cm,height=4cm,keepaspectratio=false]{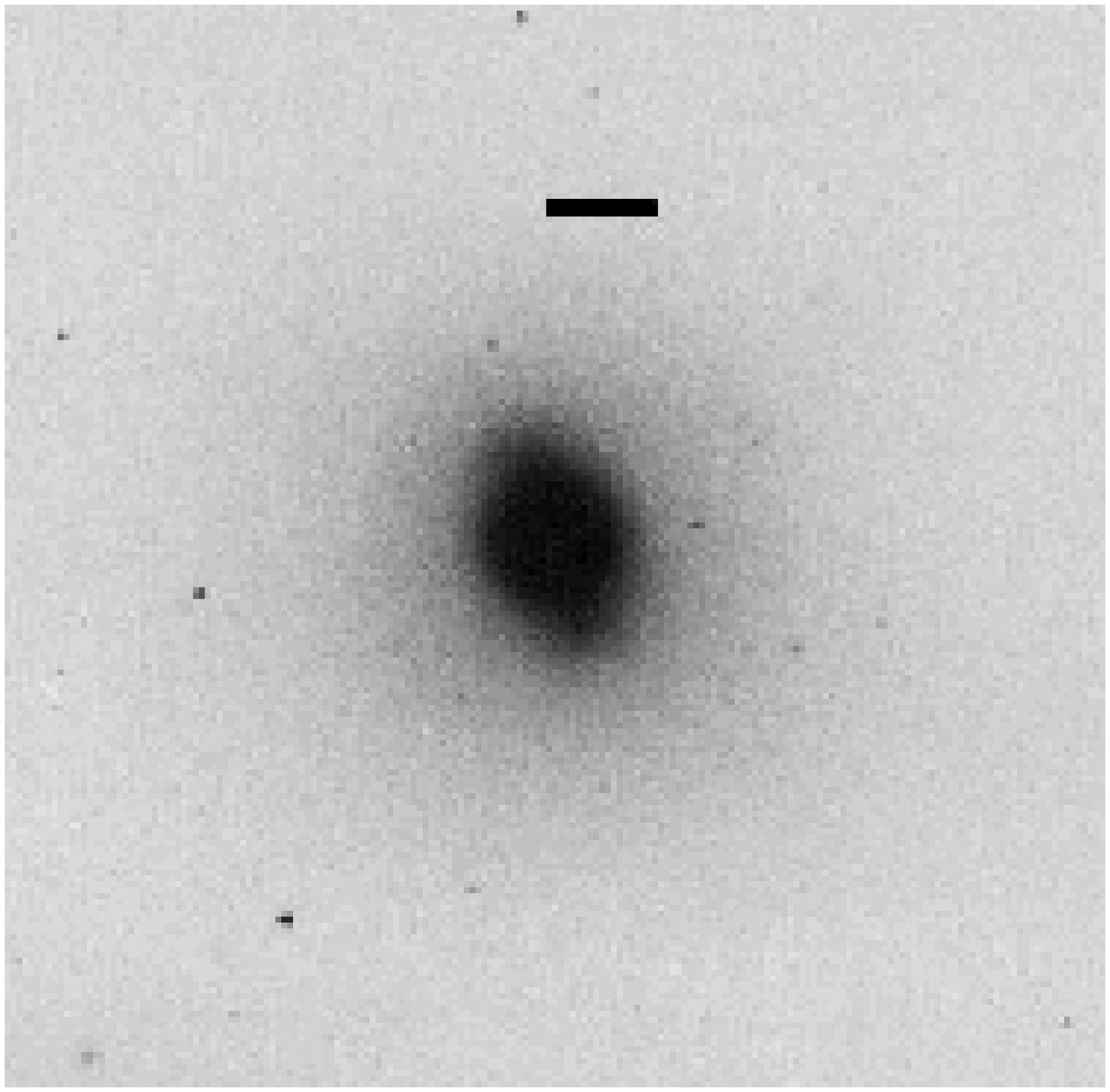}\\
\includegraphics[width=4cm,height=4cm,keepaspectratio=false]{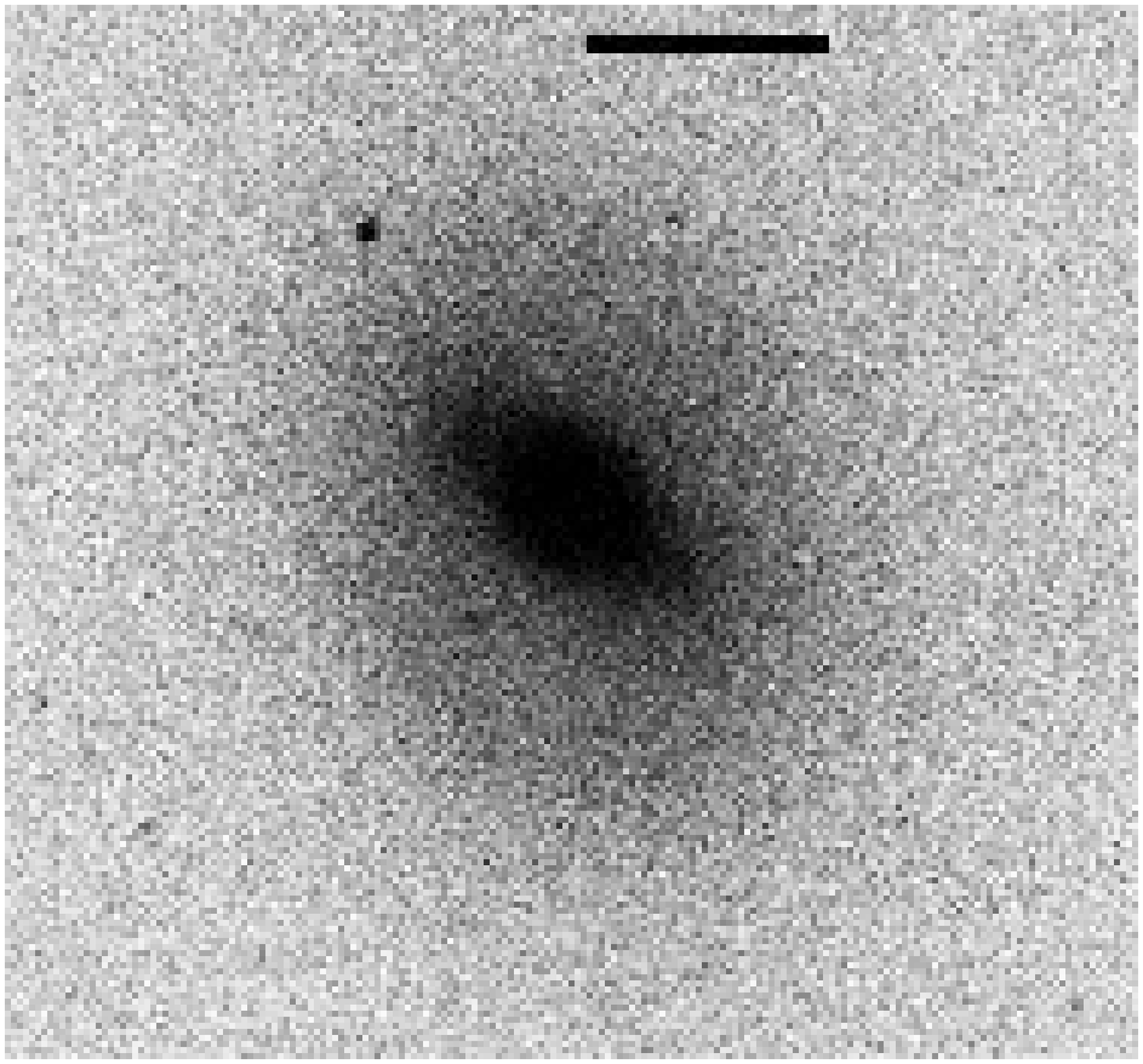}
\includegraphics[width=4cm,height=4cm,keepaspectratio=false]{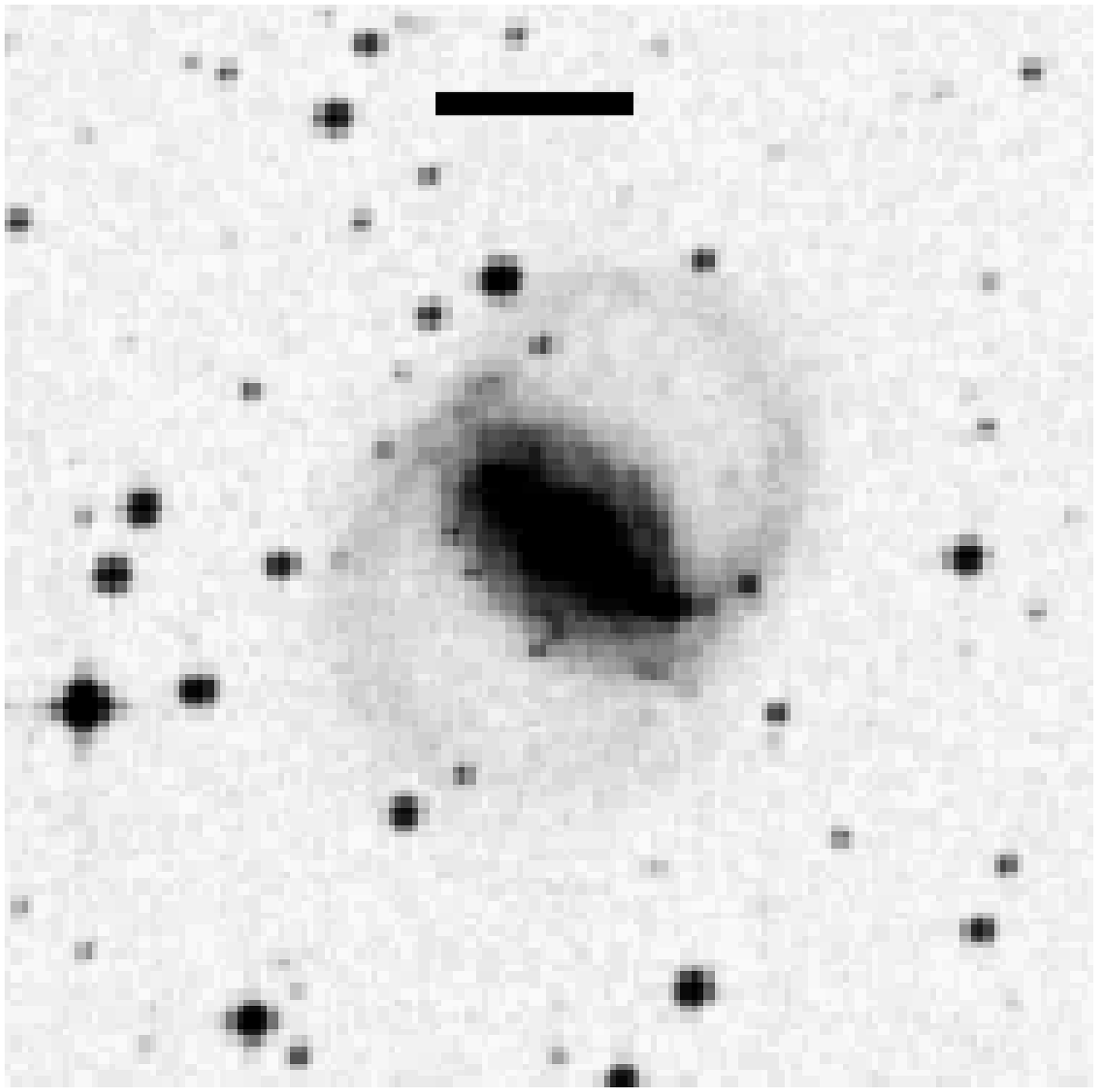}
\includegraphics[width=4cm,height=4cm,keepaspectratio=false]{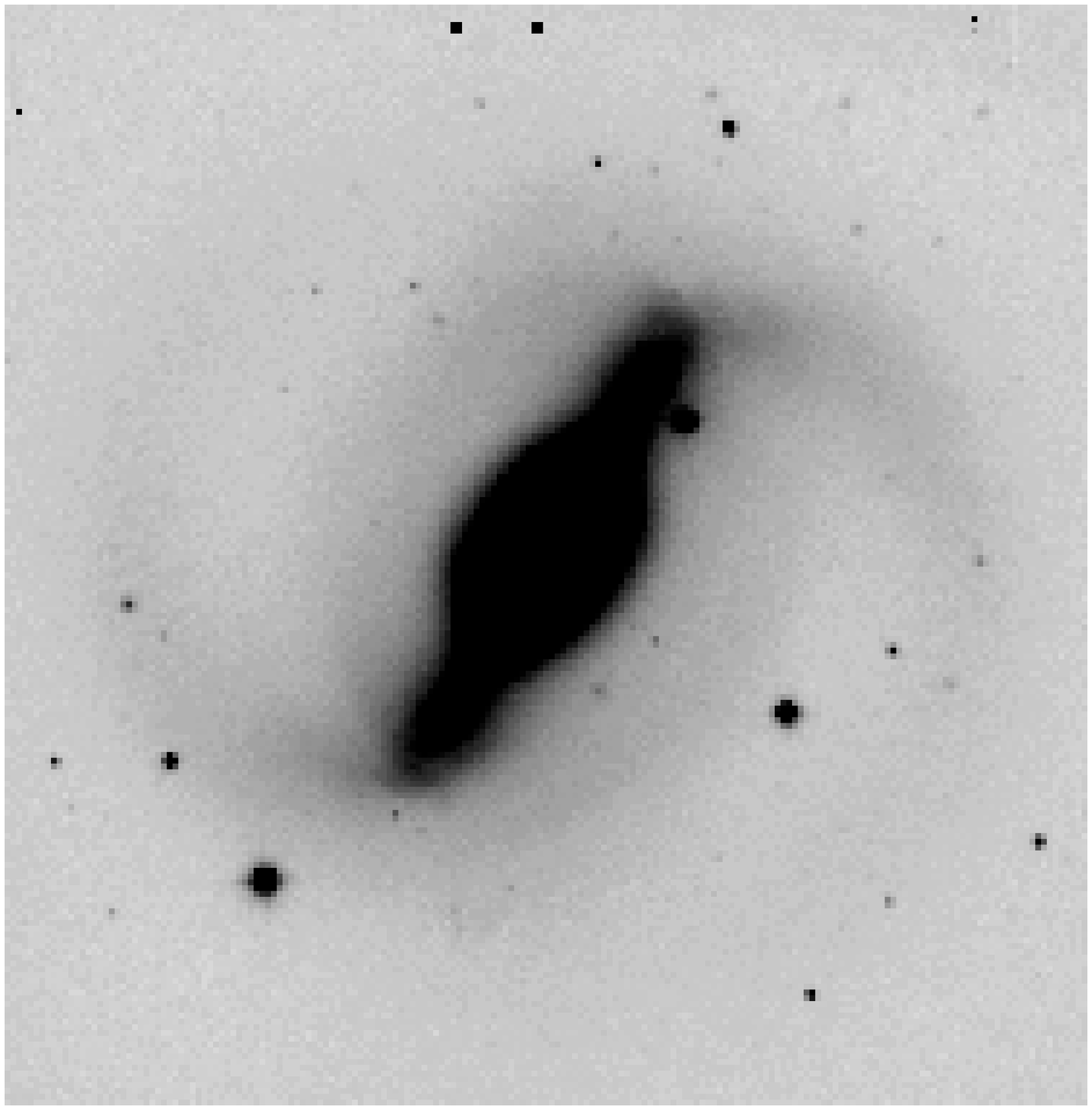}
\includegraphics[width=4cm,height=4cm,keepaspectratio=false]{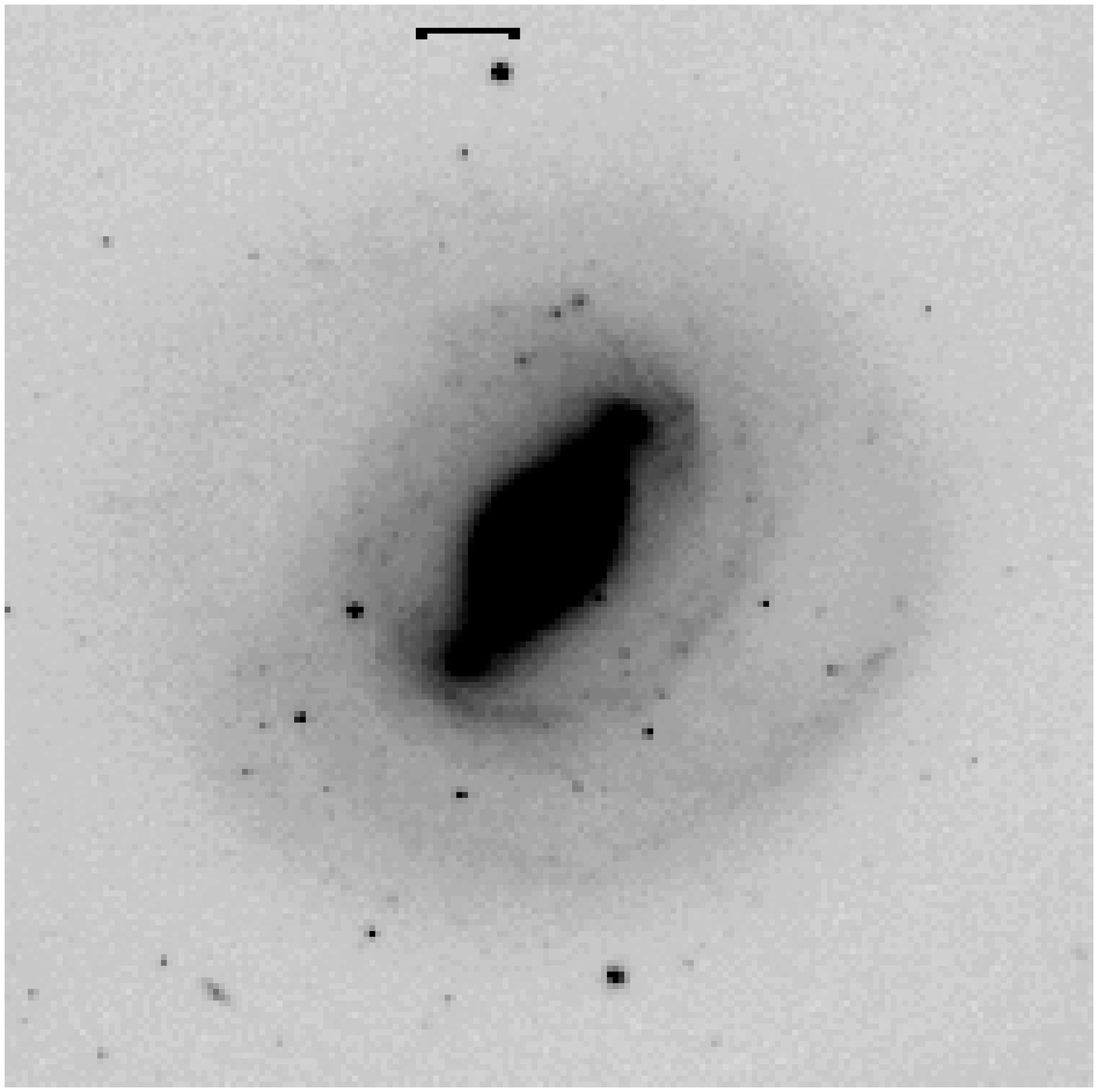}\\
\includegraphics[width=4cm,height=4cm,keepaspectratio=false]{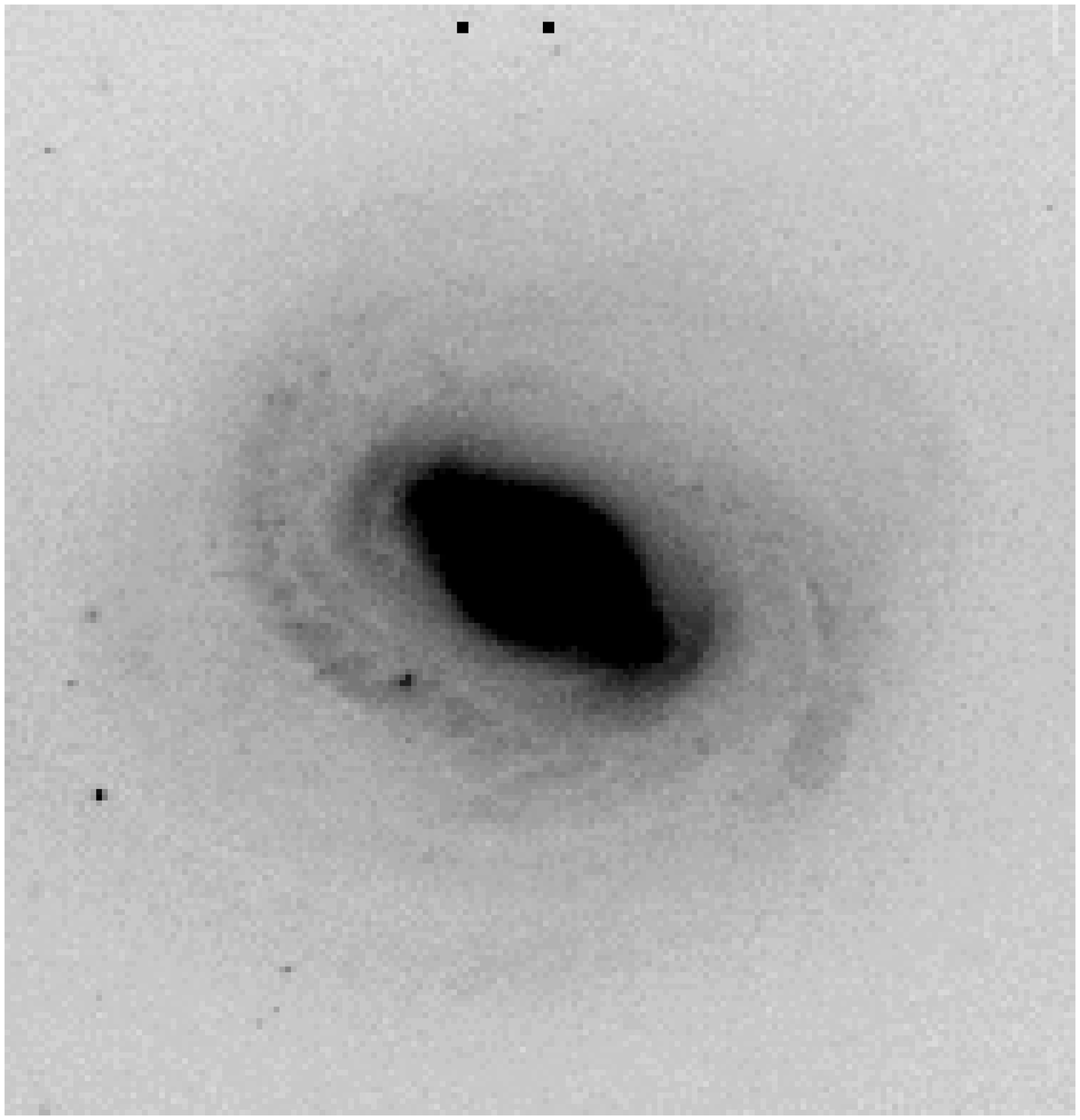}
\includegraphics[width=4cm,height=4cm,keepaspectratio=false]{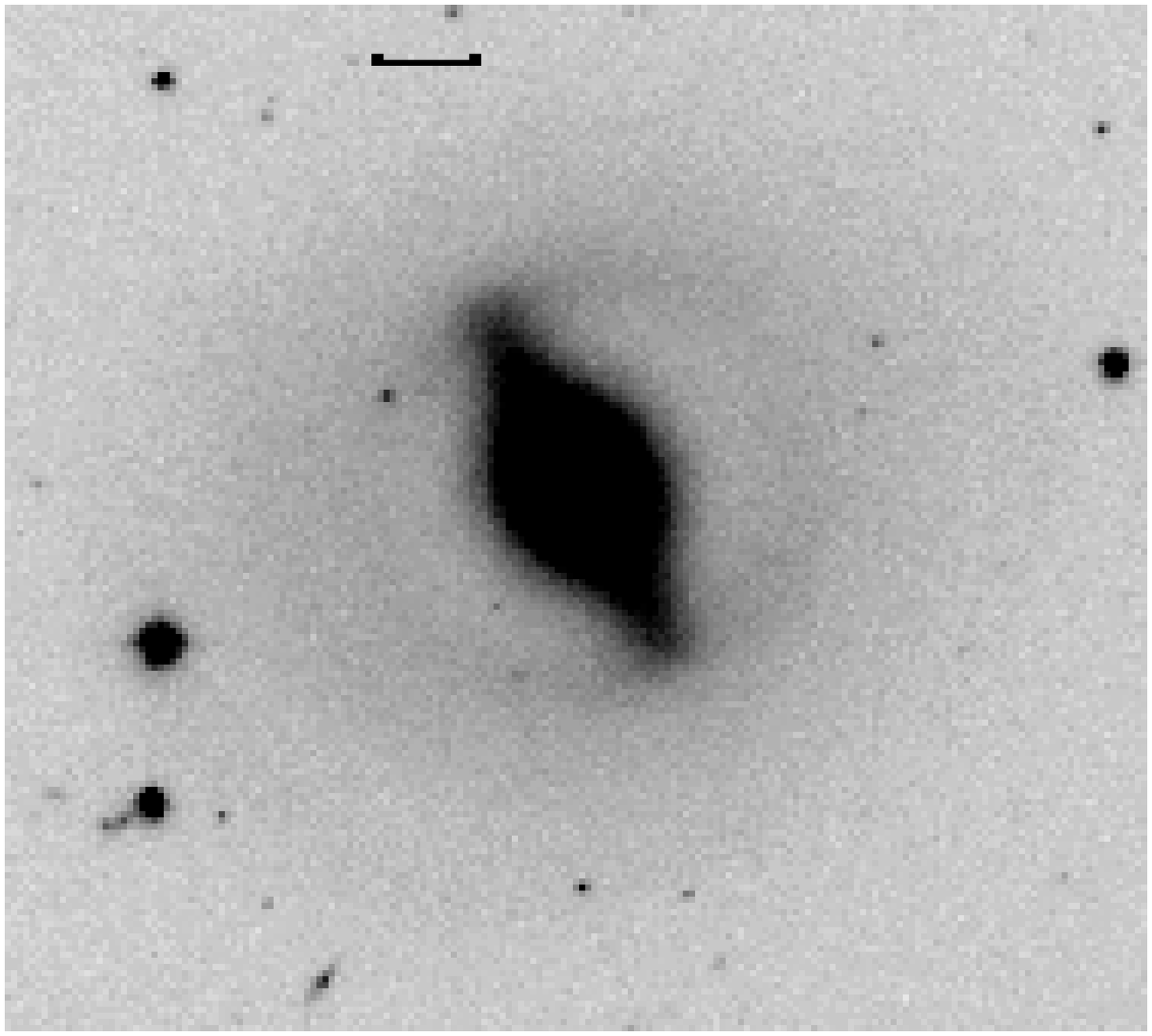}
\includegraphics[width=4cm,height=4cm,keepaspectratio=false]{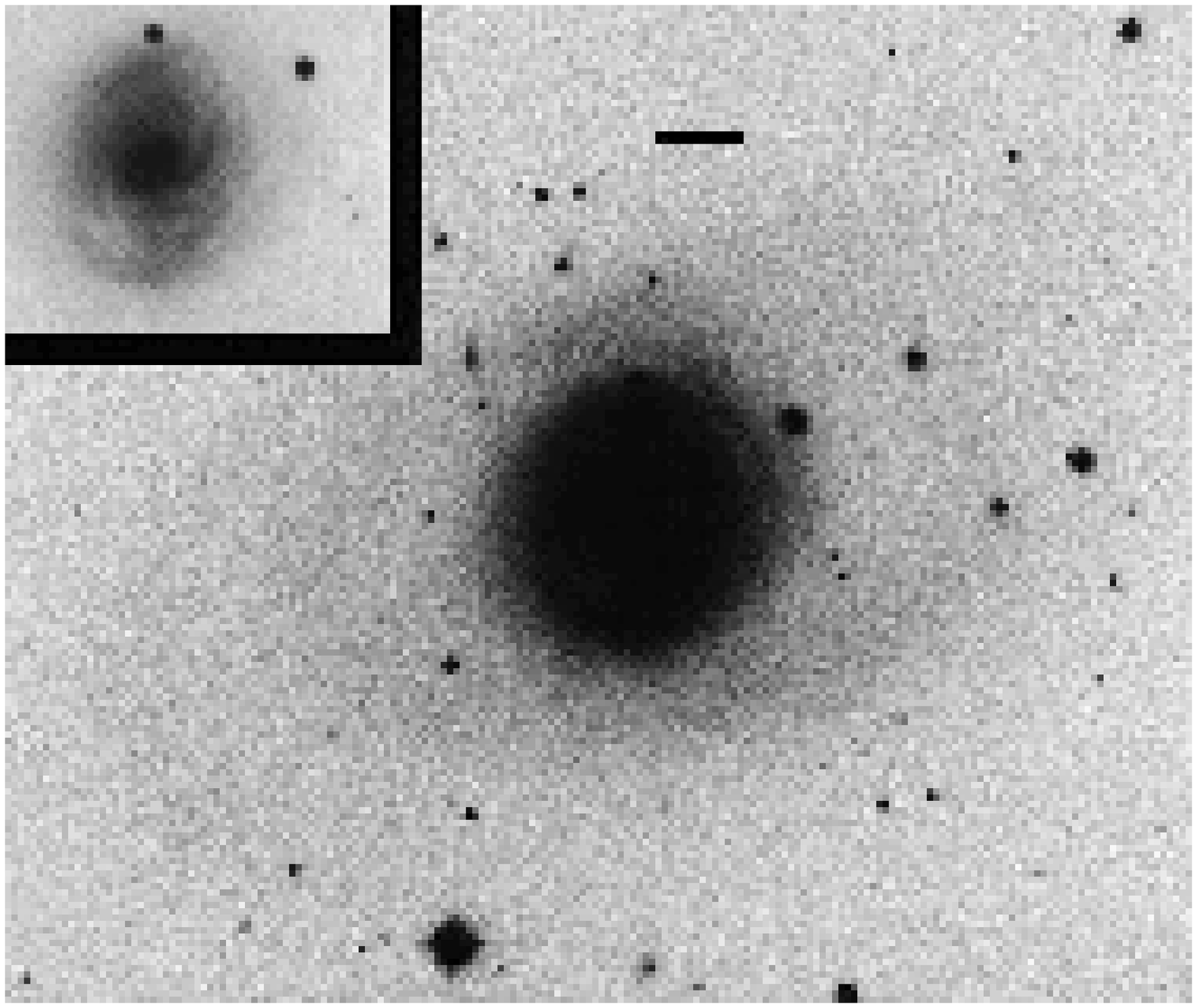}
\includegraphics[width=4cm,height=4cm,keepaspectratio=false]{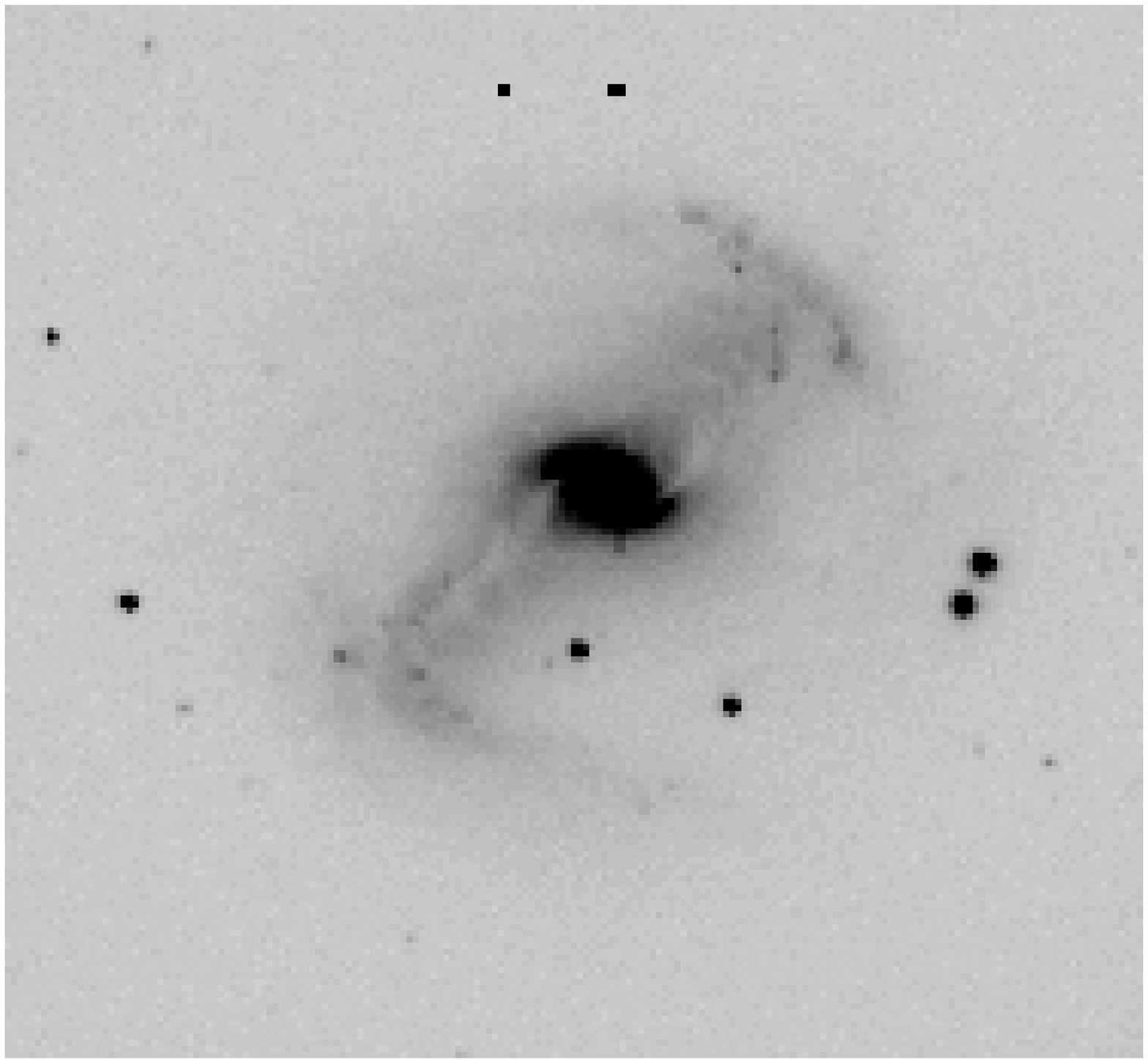}\\
\includegraphics[width=4cm,height=4cm,keepaspectratio=false]{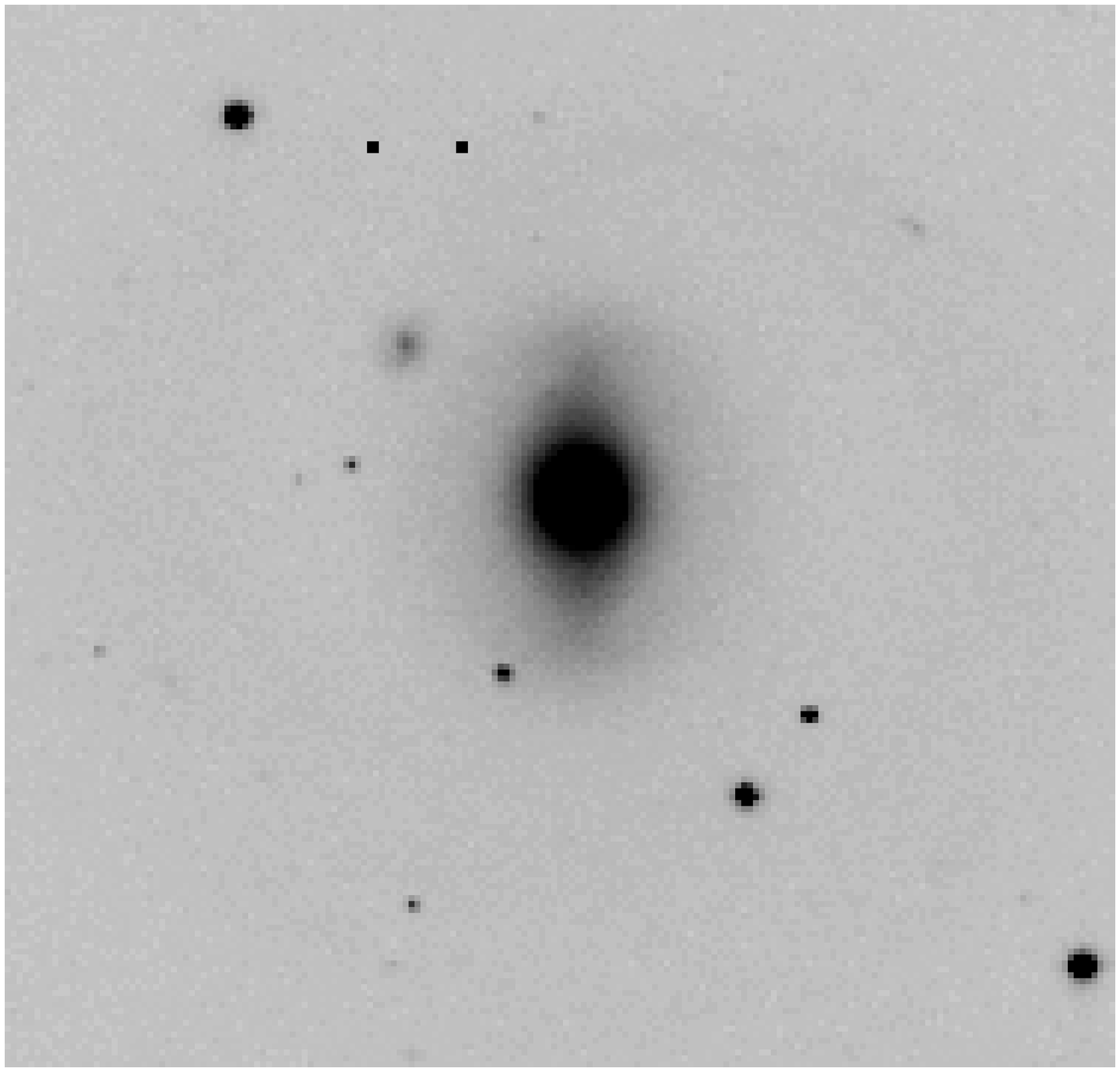}
\includegraphics[width=4cm,height=4cm,keepaspectratio=false]{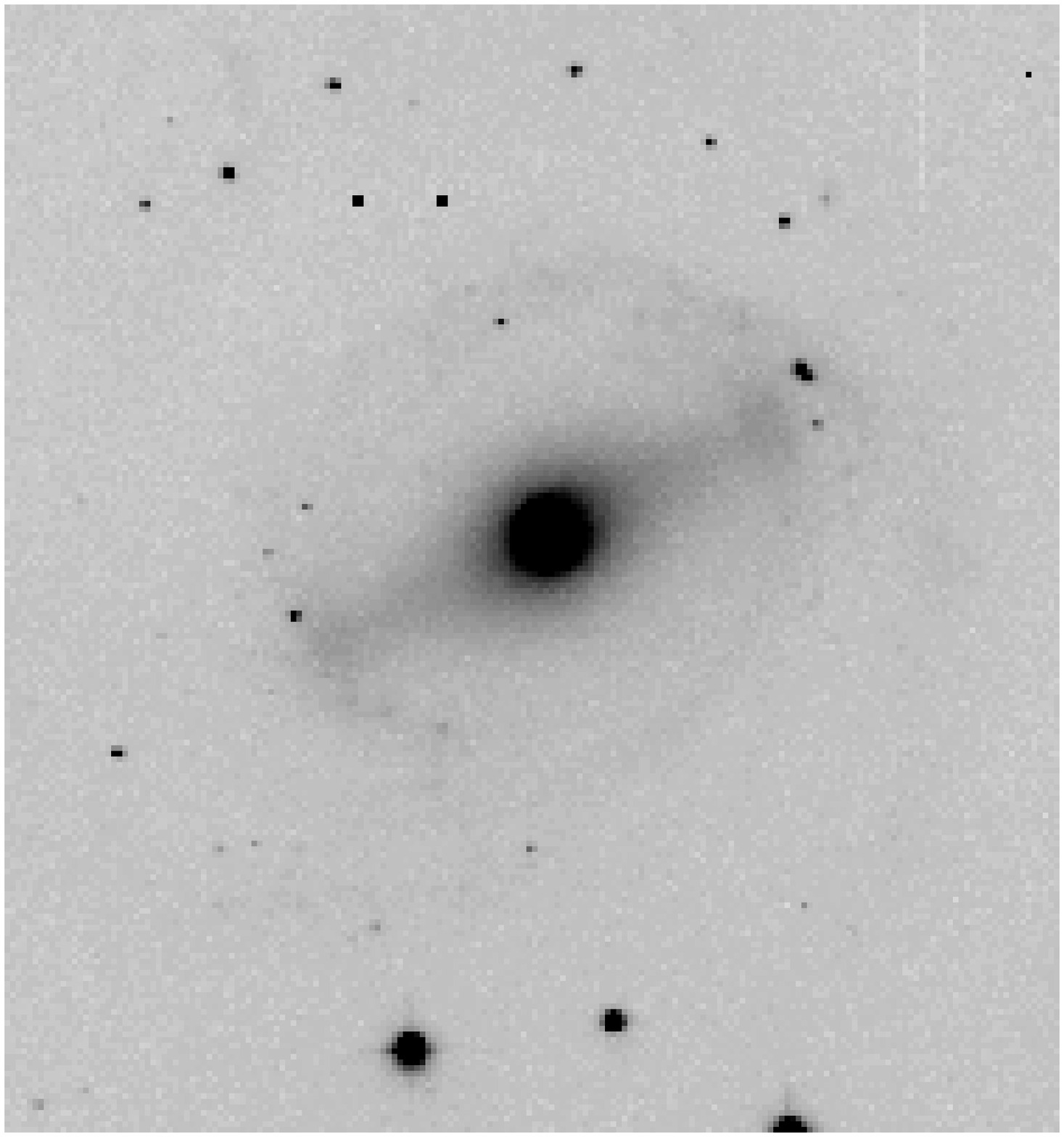}
\caption{All galaxies from our sample. From left to right and from top to bottom:
NGC 1302, 1317, 1326, 1387, 1440, 2665, 4314, 4394, 4579, 4608, 4984, 5383, 5701, and
NGC 5850. Horizontal lines in each panel have approximately a 20 arcseconds length, except
for NGC 2665, where it has 30 arcseconds. These images were taken from the Carnegie Atlas of Galaxies
\citep{san94}, the Digitized Sky Survey, and from our own R CCD images \citep{gad05}.}
\end{figure}

\clearpage

\begin{figure}
\begin{center}
\includegraphics[width=10cm,keepaspectratio=true]{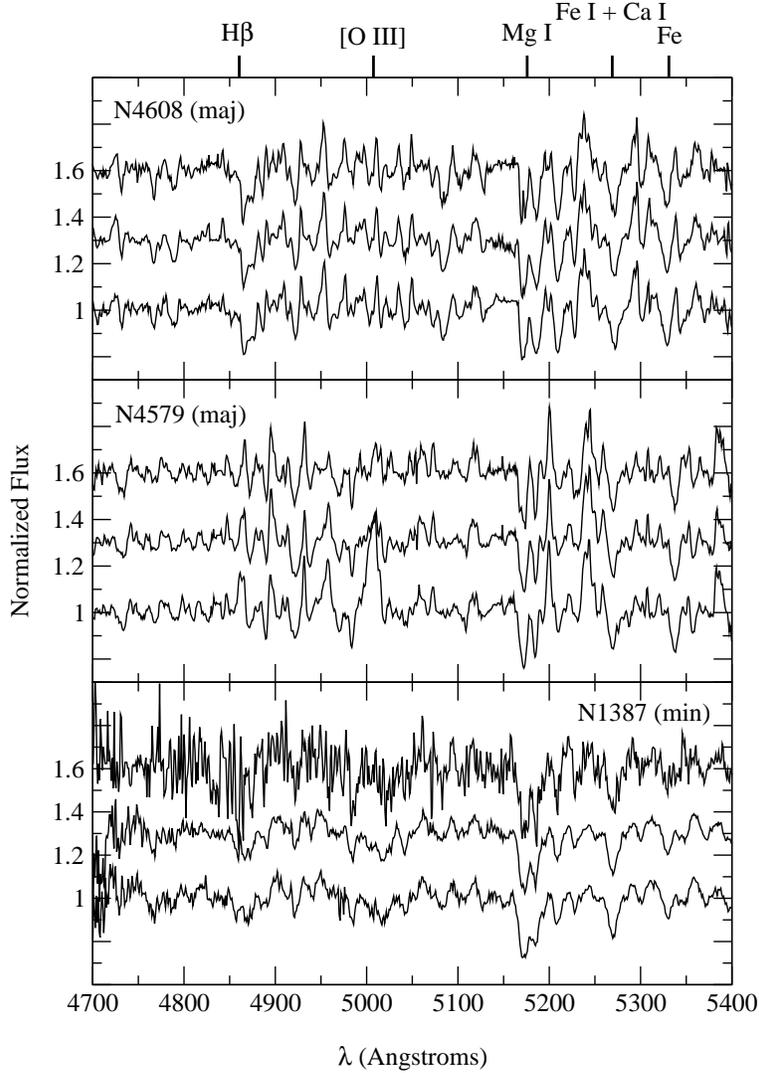}
\end{center}
\caption{Some typical examples of the spectra we have obtained. The two upper panels refer to
spectra obtained along the bar major axis of NGC 4608 and NGC 4579, both from the North sample.
The lower panel shows spectra obtained along the minor axis of the bar in NGC 1387, from the South
sample. For each galaxy, as
indicated, the lower spectrum is the central one, while the middle one was extracted at
4.5'' from the center, and the upper one at 19.3''. The latter were artificially dislocated in this figure to
avoid crowding. The emission line at $\lambda\approx5200$ {\AA} in NGC 4579 is the [N I] doublet.}
\end{figure}

\clearpage

\begin{figure}
\begin{center}
\includegraphics[width=10cm,keepaspectratio=true]{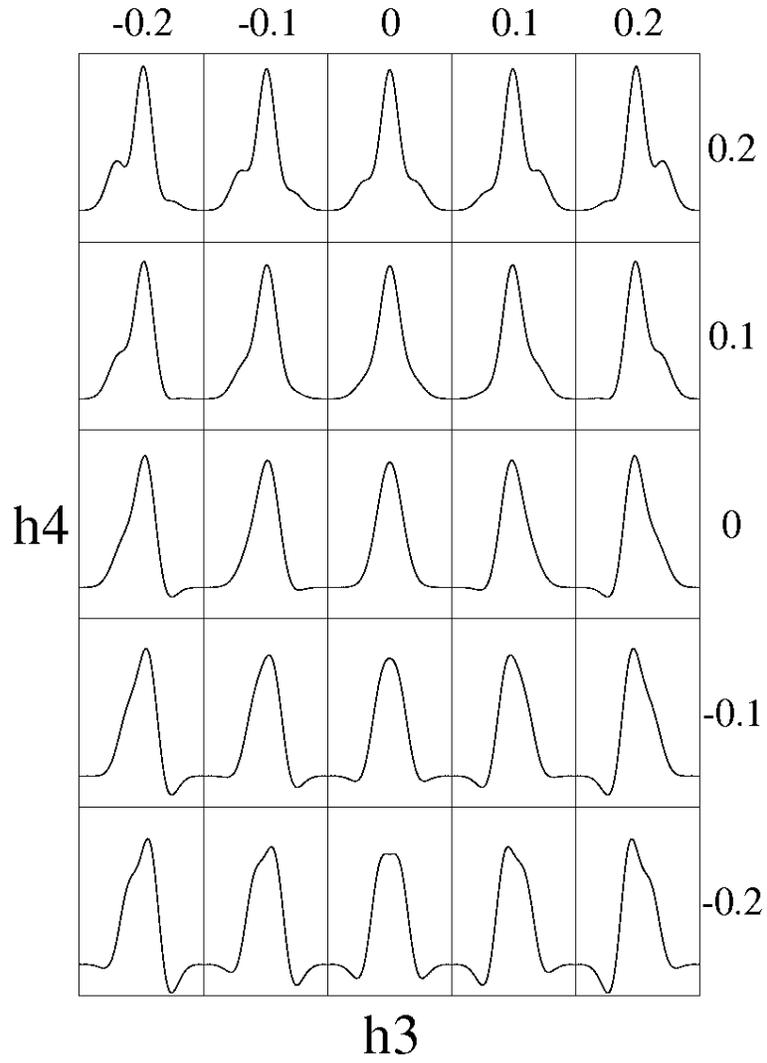}
\end{center}
\caption{A pure gaussian ($h_3=h_4=0$) may suffer asymmetric deviations when
$h_3\neq0$ and symmetric deviations when $h_4\neq0$.}
\end{figure}

\clearpage

\begin{figure}
\begin{center}
\includegraphics[width=9.5cm,keepaspectratio=true]{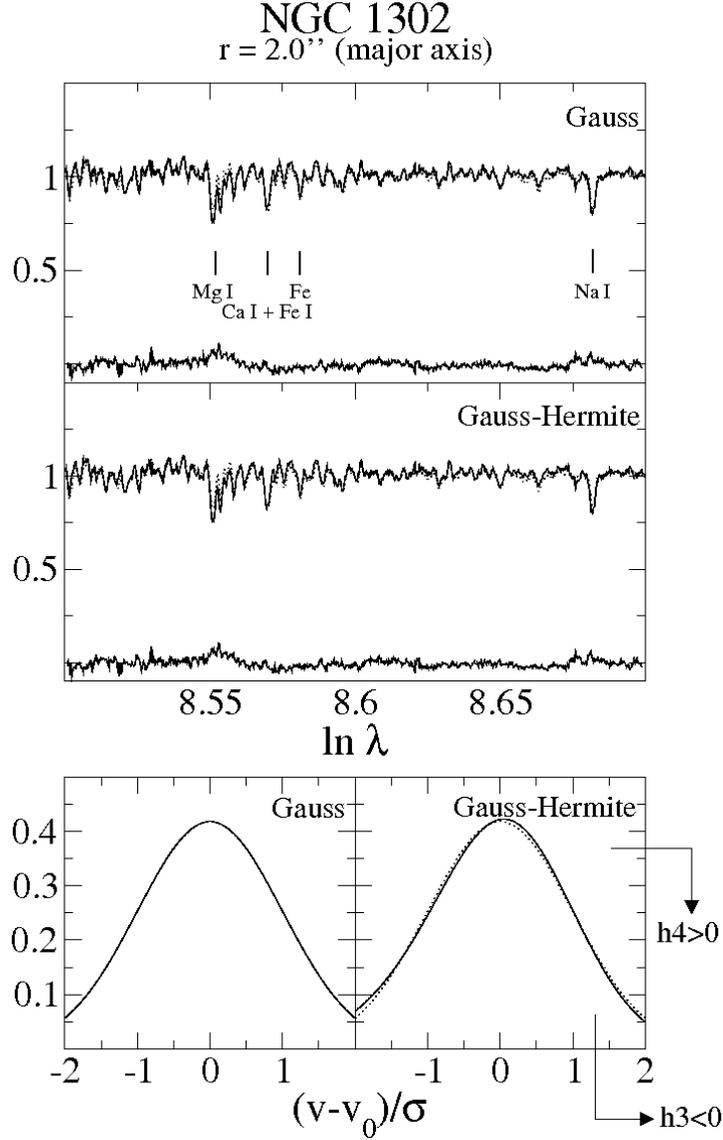}
\end{center}
\caption{An example of the fits generated by our code to retrieve the LOSVDs from the galaxy
spectra, in this case NGC 1302 at 2.0'' from the center along its bar major axis. The results
from the two parameterizations are shown: the velocity distribution as a pure gaussian (upper panel
and lower left panel), and as a generalized gaussian (Gauss--Hermite series, middle panel and
lower right panel). The upper and middle panels show the galaxy spectrum (solid line) and the
solution found with the template stars and the determined LOSVD (dotted line). Residuals are also
shown. The lower panels show the determined LOSVDs. Note the effects of the higher order
moments of the generalized gaussian. In the lower right panel the dotted line is a pure gaussian profile
for comparison.}
\end{figure}

\clearpage

\begin{figure}
\begin{center}

\includegraphics[width=10cm,keepaspectratio=true]{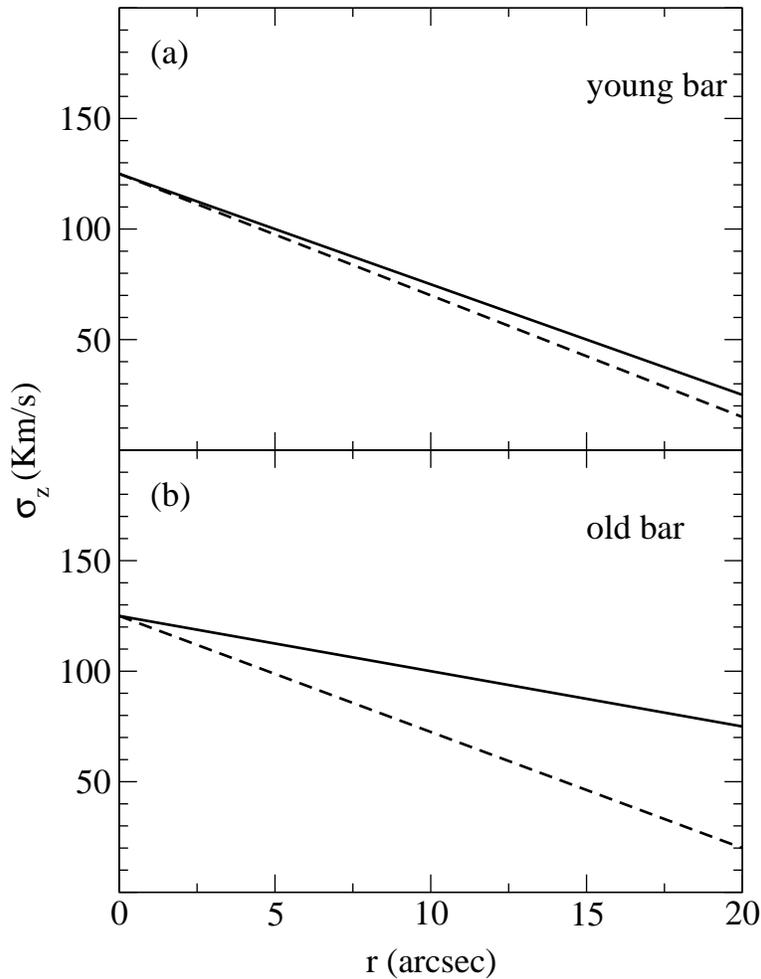}
\end{center}
\caption{If one is able to measure $\sigma_z$ not only
along the bar major axis (solid line) but also in the bar minor axis (dashed line), reaching in the
latter case regions in the galaxy dominated only by the disk, i.e., outside the contributions from
the bulge and the bar, then a simple comparison of the values of $\sigma_z$ in the bar and in the disk
will suffice to evaluate if the bar is young or old. A young bar (a) has a velocity dispersion
yet similar to the one in the disk, whereas an evolved bar (b) has a much higher $\sigma_z$ than the disk
due to its dynamical evolution.
Note that the units for the abscissae are of course only meant to represent our study. The location of
the different $\sigma_z$ behaviors will evidently vary in different galaxies.}
\end{figure}

\clearpage

\begin{figure}
\begin{center}
\includegraphics[width=16cm,keepaspectratio=true]{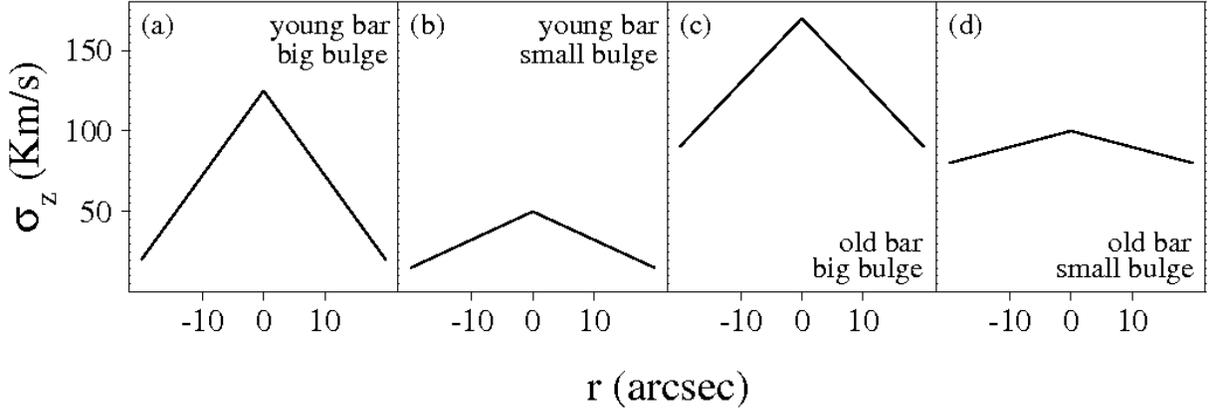}
\end{center}
\caption{Typical radial profiles of $\sigma_z$ for young and old bars in the presence of a bulge.
In most cases, the acquisition of spectra from pure disk regions is very difficult and
even the outermost spectra along the minor axis may be still within the bar. In these
cases the radial behavior of $\sigma_z$ is
somewhat similar along both the major and minor axes of the bar.
The kinematics of the bulge basically affects the central velocity dispersion.
The two leftmost panels show
examples of recently formed bars, revealed by the low values for $\sigma_z$ typical of disk stars,
while in the
rightmost panels the evolved bars may be recognized by values of $\sigma_z$ that can not be ascribed
to a disk. In (a) and (c), however, the bulge is dynamically hotter than the bar, even an evolved one.
In (b) and (d), on the other hand, the kinematics of the bulge and the bar are similar, and this may be
true for a young or for an old bar. Morphological differences may also account for this different
behaviors, as bulges of earlier--type galaxies have a larger velocity dispersion.
Note that the units for the abscissae are of course only meant to represent our study. The location of
the different $\sigma_z$ behaviors will evidently vary in different galaxies.}
\end{figure}

\clearpage

\begin{figure}
\begin{center}
\includegraphics[width=11cm,keepaspectratio=true]{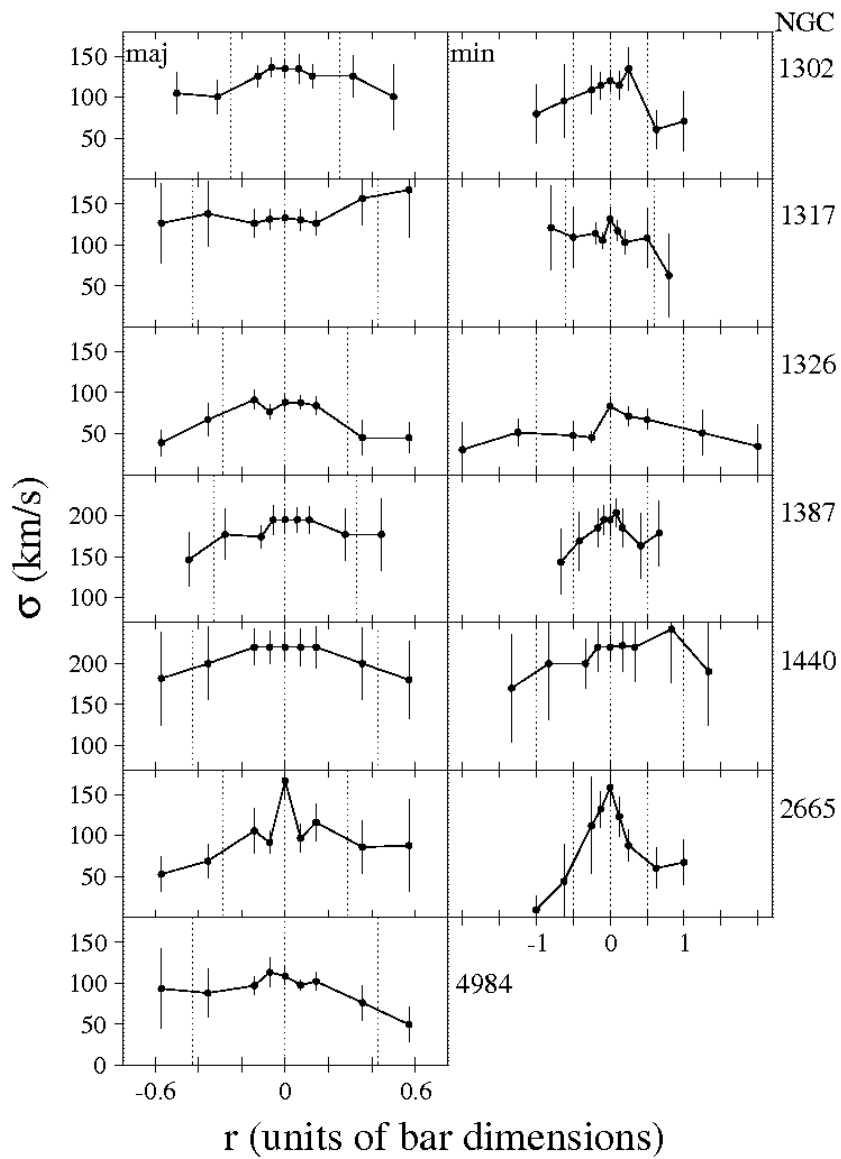}
\end{center}
\caption{Vertical velocity dispersion radial profiles for the galaxies in our South sample
along the major and minor axes of the bars, as displayed. The figures
refer to a parameterization of the LOSVDs as a pure gaussian, but similar results were obtained
when we made use also of the $h_3$ and $h_4$ higher order moments of the Gauss--Hermite series.
Dotted lines mark the center and the region where the emitted light is dominated
by the bulge. Units in the abscissae are normalized by the bar semi-major and semi-minor axes.}
\end{figure}

\clearpage

\begin{figure}
\begin{center}
\includegraphics[width=11cm,keepaspectratio=true]{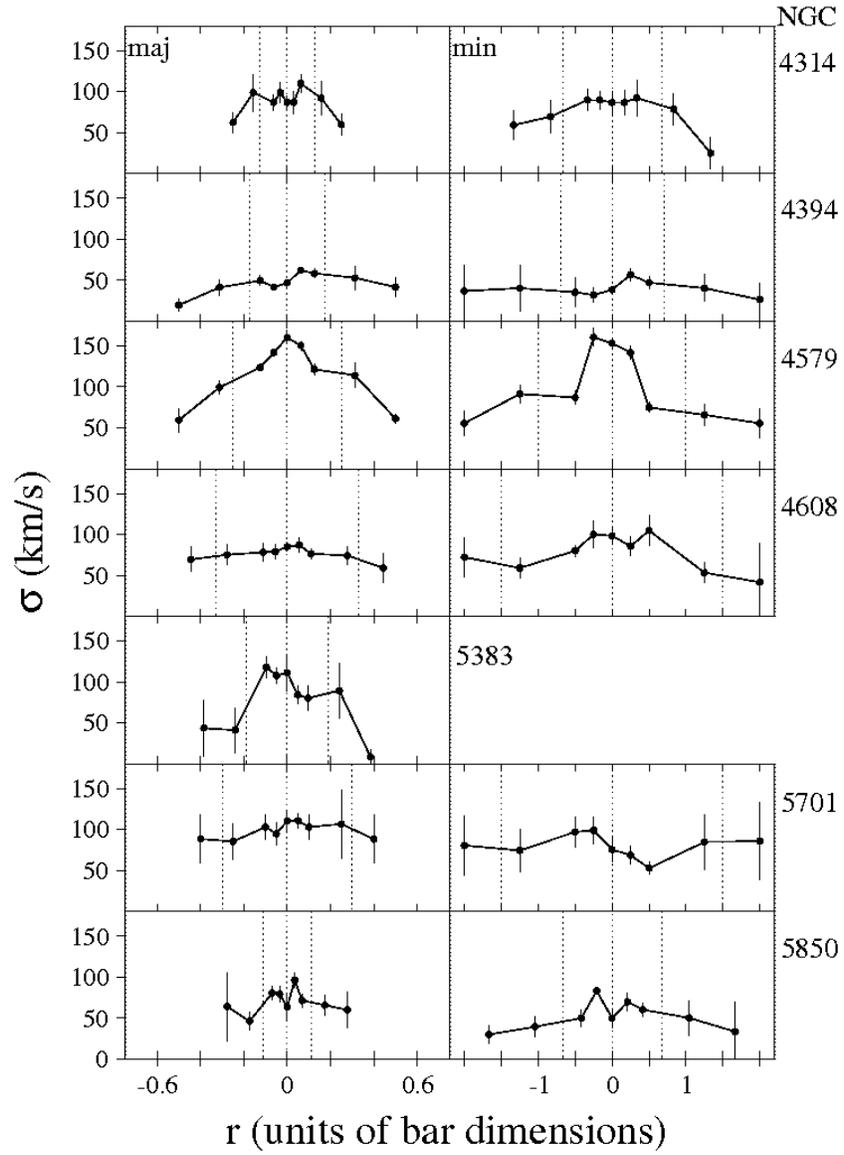}
\end{center}
\caption{Same as Fig. 7 but for the North sample.}
\end{figure}

\clearpage

\begin{figure}
\begin{center}
\includegraphics[width=8cm,keepaspectratio=true]{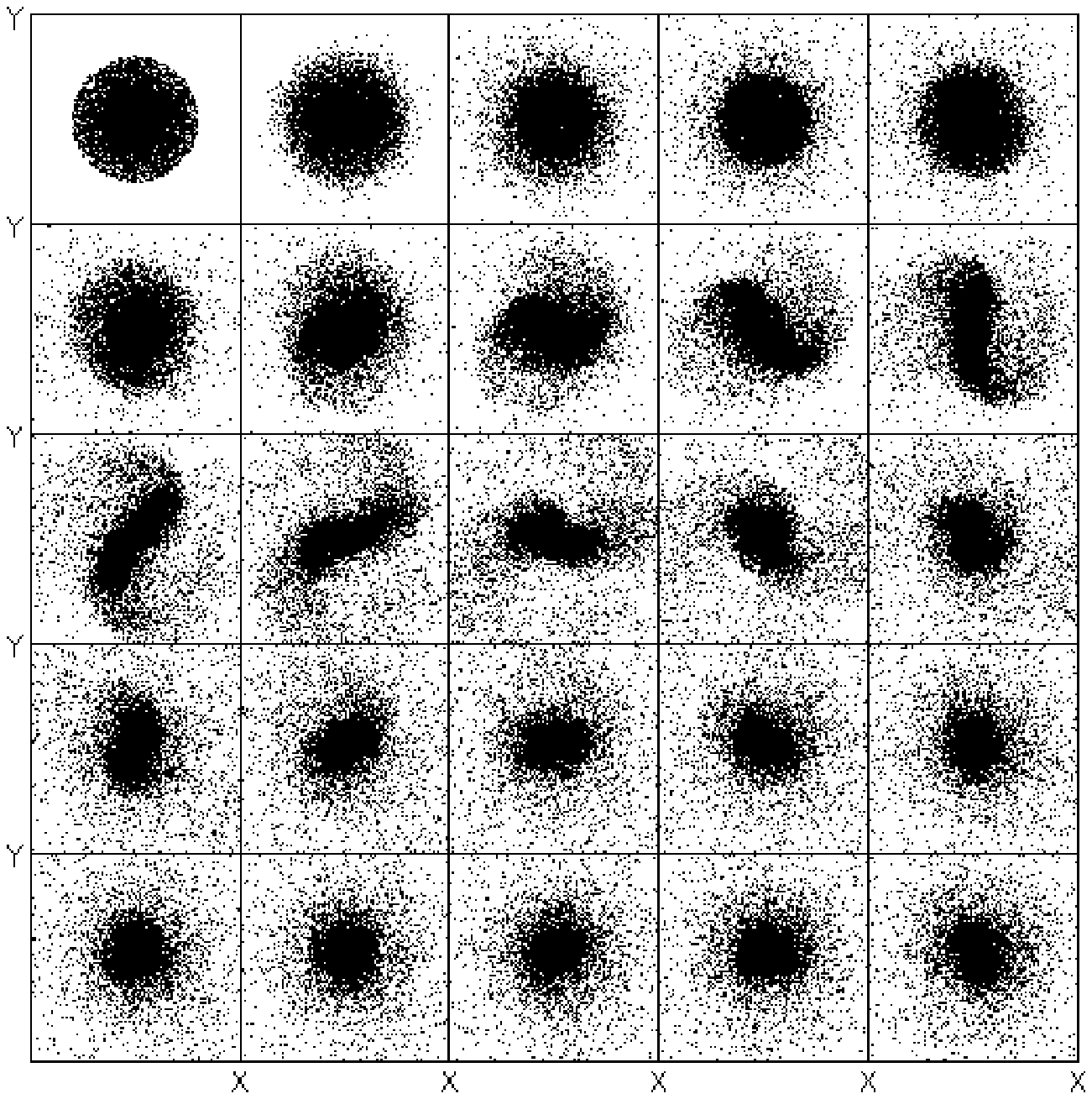}
\includegraphics[width=8cm,keepaspectratio=true]{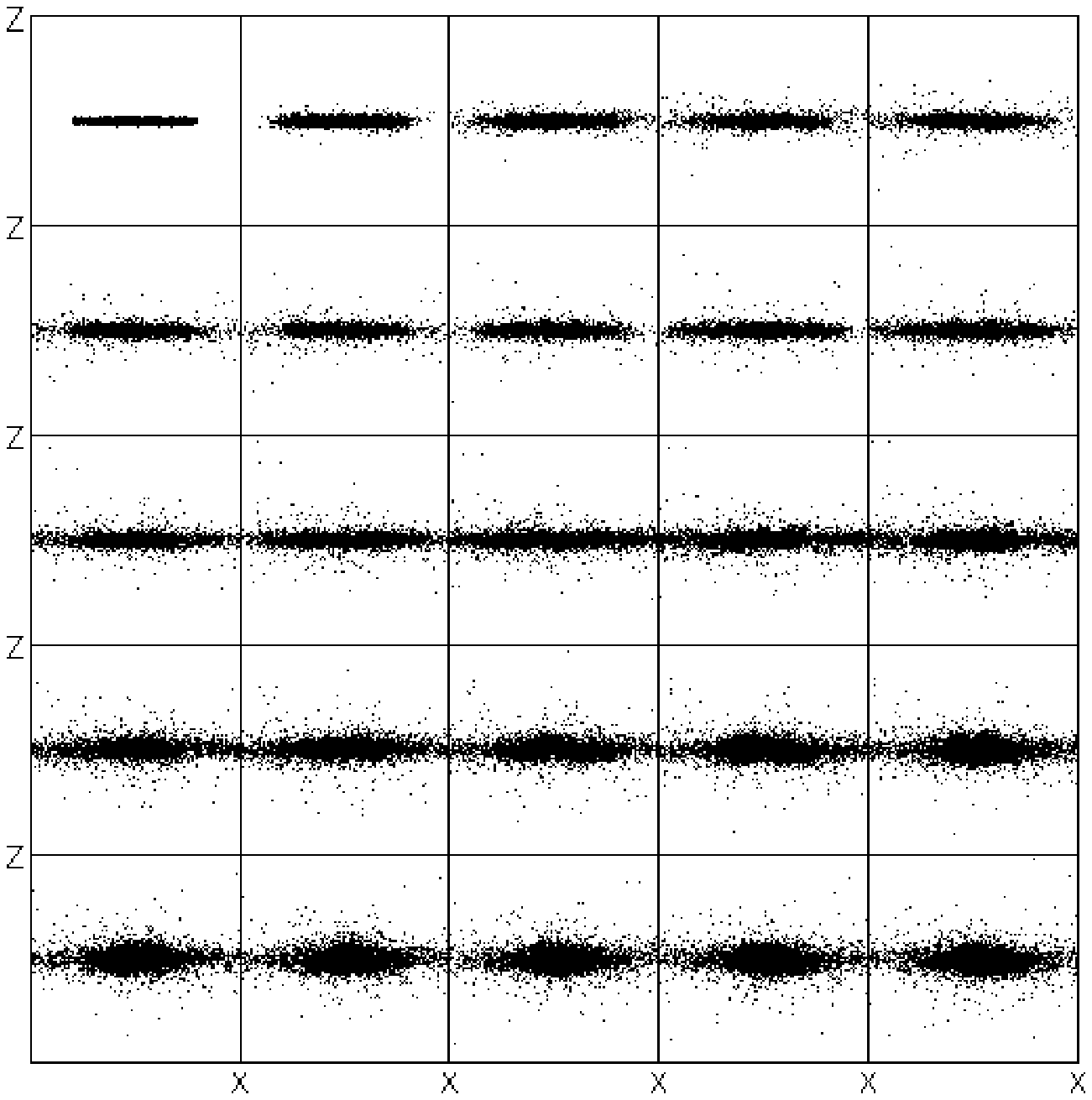}
\end{center}
\caption{Formation of the grand design morphology in the evolution of a pure stellar disk seen
face--on (left) and edge--on (right). The top leftmost panels show the initial conditions, whereas
the bottom rightmost panels refer to $t=1.9$ Gyr of evolution. The time interval between each panel
is $t=8\times10^7$ yr and its physical dimension is 16 Kpc. Only 10\% of the particles in this
simulation are shown.}
\end{figure}

\clearpage

\begin{figure}
\begin{center}
\includegraphics[width=16cm,keepaspectratio=true]{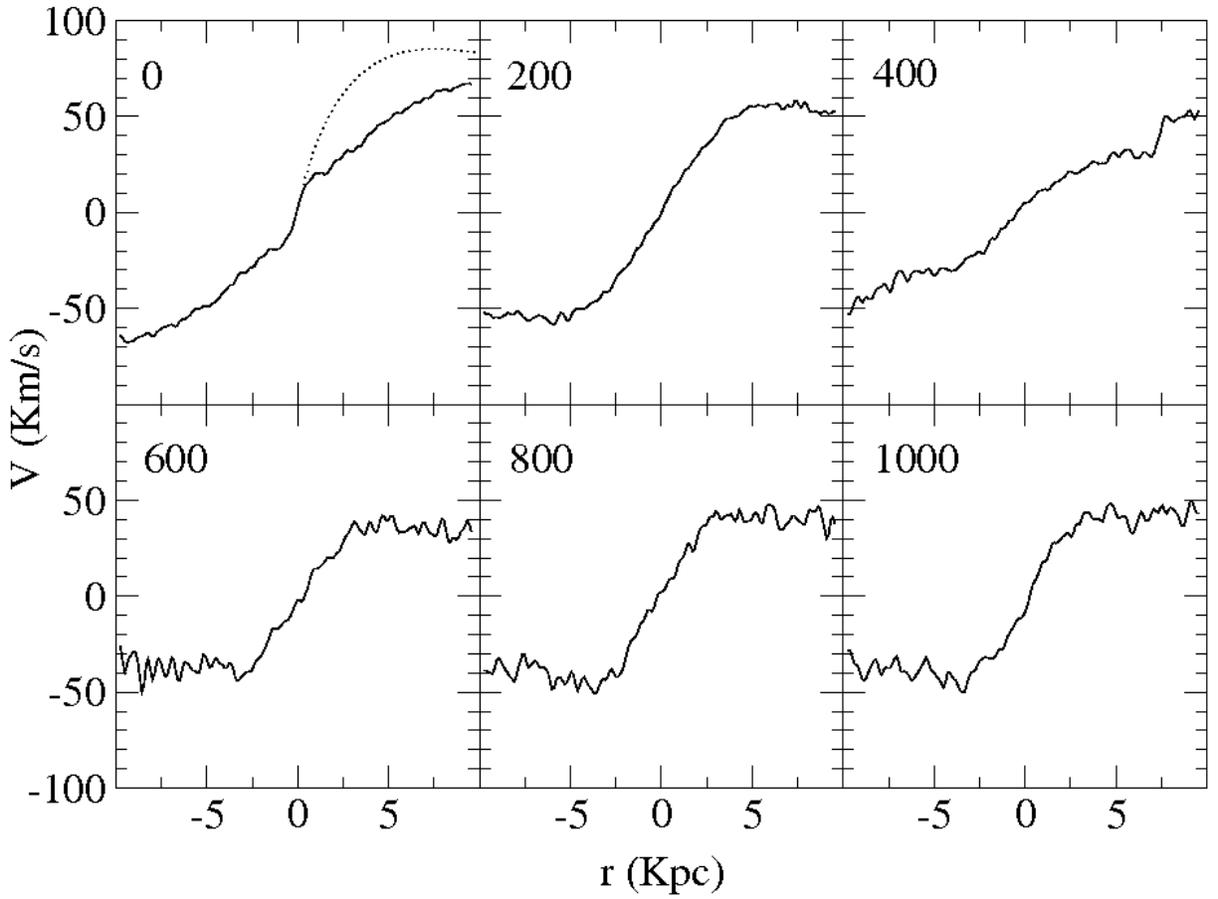}
\end{center}
\caption{Evolution in time of the rotation curve, as measured from simulated long slit spectroscopy,
in our representative stellar disk numerical simulation.
The corresponding times in virial units are displayed
in the top left of each panel. The last panel corresponds to 2 Gyr of evolution. The dotted line in the
first panel is the initial circular speed curve, for comparison.}
\end{figure}

\clearpage

\begin{figure}
\begin{center}
\includegraphics[width=6cm,keepaspectratio=true]{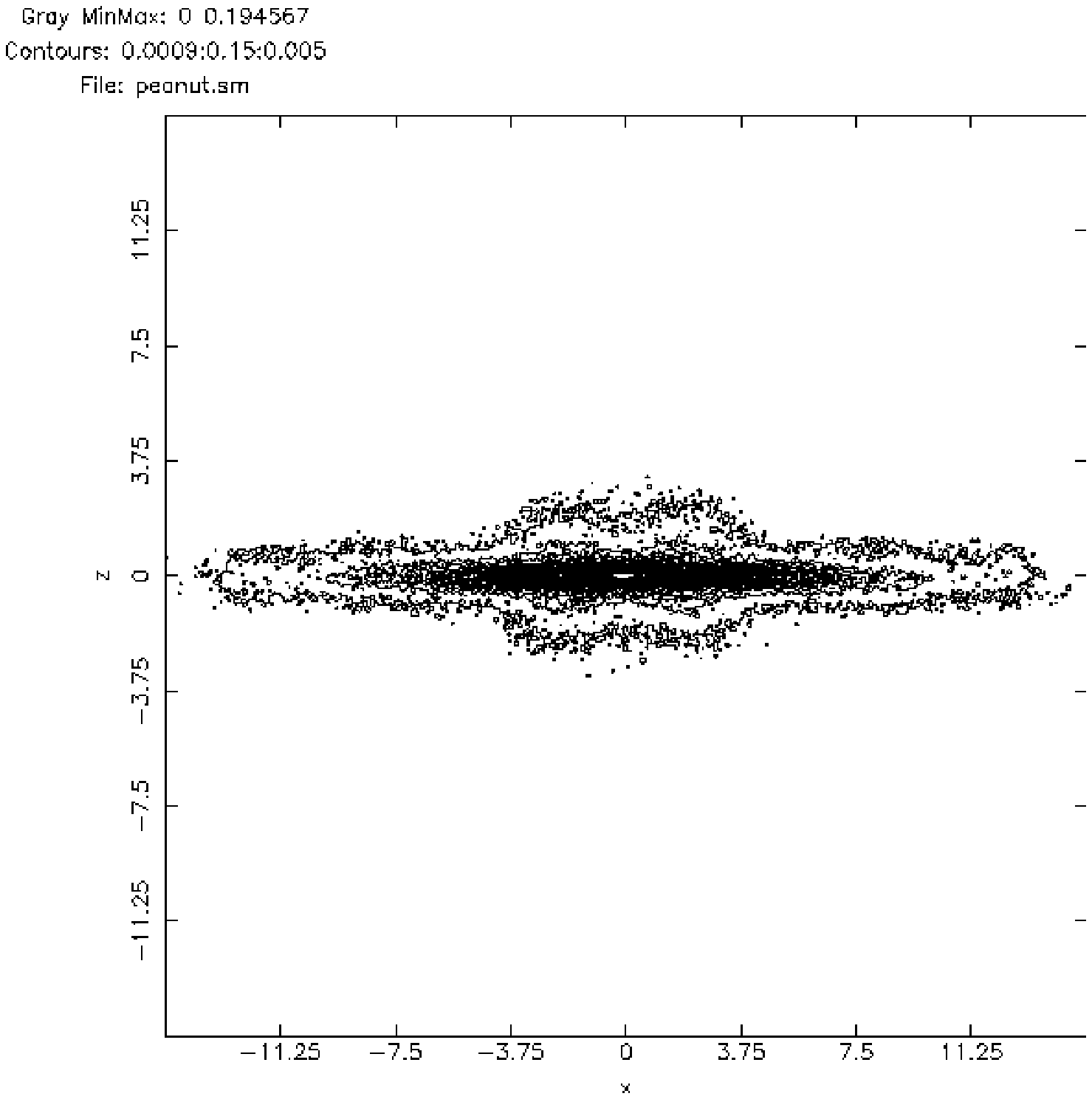}
\includegraphics[width=6cm,keepaspectratio=true]{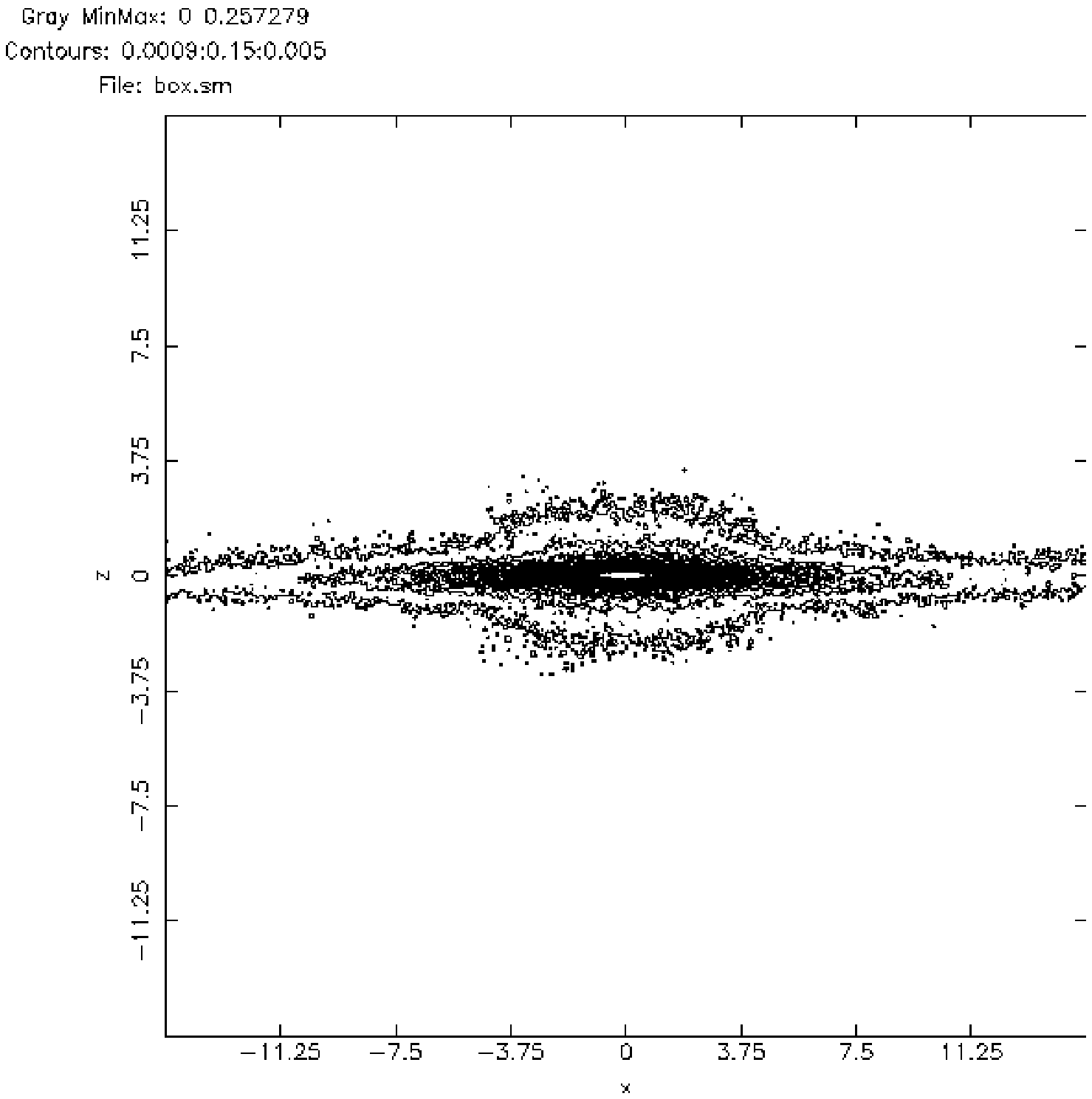}
\end{center}
\caption{Isodensity contours in our fiducial numerical experiment seen edge--on
in $t=3.6\times10^8$ yr (left) and in $t=4\times10^8$ yr (right).
Note that the peanut morphology appears when the line of sight is closer to a
perpendicular orientation with respect to the bar major axis (left), whereas the boxy morphology
appears when the line of sight is closer to a parallel orientation with respect to the bar major axis
(right). Dimension units are 650 pc.}
\end{figure}

\clearpage

\begin{figure}
\begin{center}
\includegraphics[width=16cm,keepaspectratio=true]{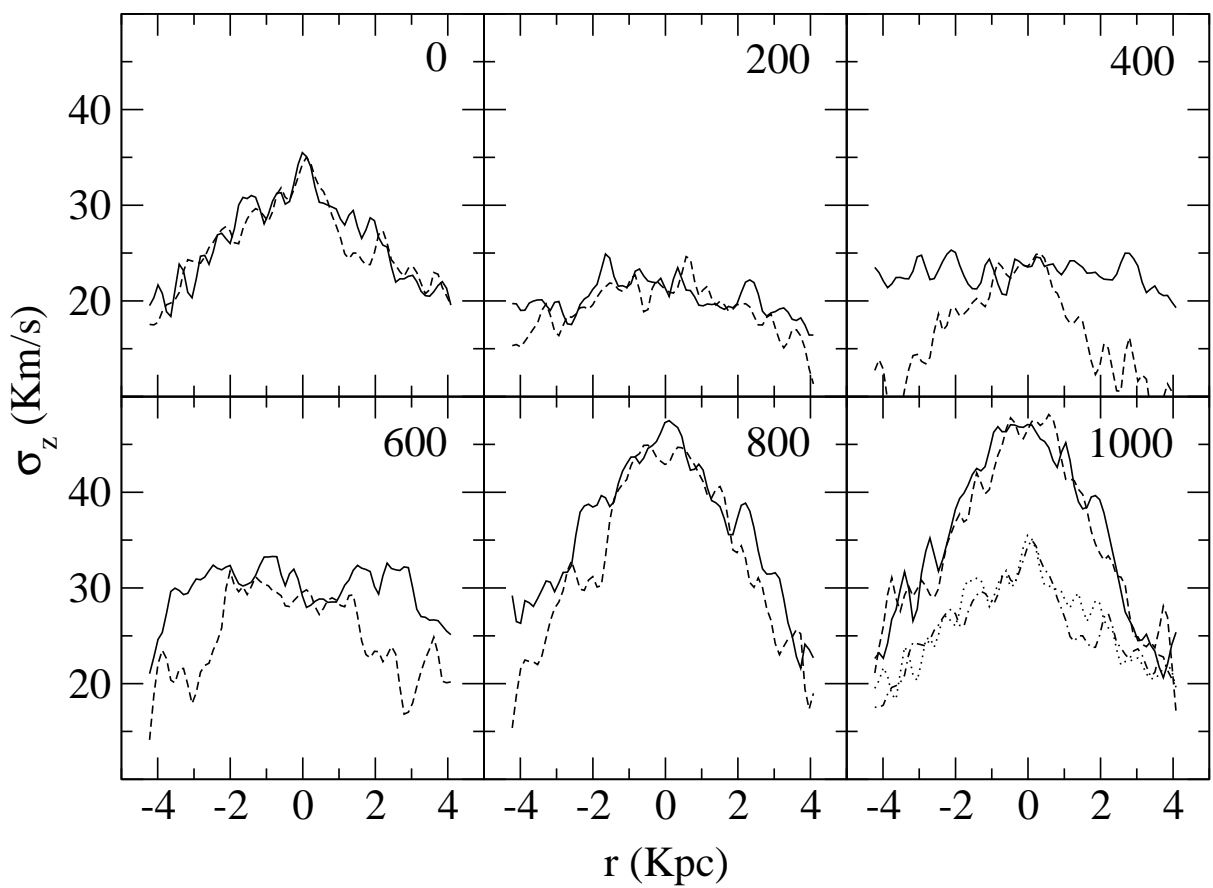}
\end{center}
\caption{Radial profiles of the vertical velocity dispersion along the bar major axis
(solid line) and minor axis (dashed line),
as measured from simulated long slit spectroscopy,
in our representative stellar disk numerical simulation.
The corresponding times in virial units are displayed
in the top right of each panel. The last panel corresponds to 2 Gyr of evolution. In the last panel
we plot the initial profiles again, as a dotted line for the major axis and as a dash--dotted
line for the minor axis, for comparison.}
\end{figure}


\begin{thebibliography}{}
\bibitem[Abramowitz \& Stegun(1965)]{abr65} Abramowitz, M., \& Stegun, I. A. 1965, Handbook of
Mathematical Functions with Formulas, Graphs, and Mathematical Tables, Dover Books on Advanced
Mathematics
\bibitem[Abt \& Biggs(1972)]{abt72} Abt, H. A., \& Biggs, E. S. 1972, Bibliography of Stellar Radial
Velocities (New York: Latham)
\bibitem[Athanassoula(1992a)]{ath92a} Athanassoula, E. 1992a, \mnras, 259, 328
\bibitem[Athanassoula(1992b)]{ath92b} Athanassoula, E. 1992b, \mnras, 259, 345
\bibitem[Athanassoula(2002)]{ath02b} Athanassoula, E. 2002, \apjl, 569, L83
\bibitem[Athanassoula(2003)]{ath03} Athanassoula, E. 2003, \mnras, 341, 1179
\bibitem[Athanassoula \& Bureau(1999)]{ath99} Athanassoula, E., \& Bureau, M. 1999, \apj, 522, 699
\bibitem[Athanassoula et al.(2000)]{ath00} Athanassoula, E., Fady, E., Lambert, J. C., \& Bosma, A.
2000, \mnras, 314, 475
\bibitem[Athanassoula \& Misiriotis(2002)]{ath02a} Athanassoula, E., \& Misiriotis, A. 2002,
\mnras, 330, 35
\bibitem[Athanassoula \& Sellwood(1986)]{ath86} Athanassoula, E., \& Sellwood, J. A. 1986,
\mnras, 221, 213
\bibitem[Balcells \& Peletier(1994)]{bal94} Balcells, M., \& Peletier, R. F. 1994, \aj, 107, 135
\bibitem[Barnes, Sellwood, \& Kosowsky(2004)]{bar04} Barnes, E. I., Sellwood, J. A., \& Kosowsky, A.
2004, \aj, 128, 2724
\bibitem[Barnes \& Hut(1986)]{bar86} Barnes, J., \& Hut, P. 1986, \nat, 324, 446
\bibitem[Berentzen et al.(1998)]{ber98} Berentzen, I., Heller, C. H., Shlosman, I., \& Fricke, K. J.
1998, \mnras, 300, 49
\bibitem[Binney(2001)]{bin01} Binney, J. 2001, ASP Conf. S., 230, 63
\bibitem[Binney \& Merrifield(1998)]{bin98} Binney, J., \& Merrifield, M. 1998, Galactic Astronomy,
(Princeton: Princeton University Press)
\bibitem[Binney \& Tremaine(1987)]{bin87} Binney, J., \& Tremaine, S. 1987, Galactic Dynamics,
(Princeton: Princeton University Press)
\bibitem[Blitz \& Spergel(1991)]{bli91} Blitz, L., \& Spergel, D. N. 1991, \apj, 379, 631
\bibitem[Bonnarel et al.(2000)]{bon00} Bonnarel, F., Fernique, P., Bienaym\'e, O., Egret, D., Genova, F.,
Louys, M., Ochsenbein, F., Wenger, M., \& Bartlett, J. G. 2000, \aaps, 143, 33
\bibitem[Bottema(1993)]{bot93} Bottema, R. 1993, \aap, 275, 16
\bibitem[Bournaud \& Combes(2002)]{bou02} Bournaud, F., \& Combes, F. 2002, \aap, 392, 83
\bibitem[Bureau \& Athanassoula(1999)]{bur99} Bureau, M., \& Athanassoula, E. 1999, \apj, 522, 686
\bibitem[Bureau \& Athanassoula(2004)]{bur04} Bureau, M., \& Athanassoula, E. 2004, \apj, submitted
\bibitem[Bureau \& Freeman(1999)]{burf99} Bureau, M., \& Freeman, K. C. 1999, \aj, 118, 126
\bibitem[Buta(1986)]{but86} Buta, R. 1986, \apjs, 61, 609
\bibitem[Buta \& Combes(1996)]{but96} Buta, R., \& Combes, F. 1996, \fcp, 17, 95
\bibitem[Carollo et al.(1997)]{car97} Carollo, C. M., Stiavelli, M., de Zeeuw, P. T., \& Mack, J. 1997,
\aj, 114, 2366
\bibitem[Cayrel de Strobel, Soubiran \& Ralite(2001)]{str01} Cayrel de Strobel G., Soubiran C., \&
Ralite N. 2001, \aap, 373, 159
\bibitem[Chung \& Bureau(2004)]{chu04} Chung, A, \& Bureau, M. 2004, \aj, 127, 3192
\bibitem[Combes \& Elmegreen(1993)]{com93} Combes, F., \& Elmegreen, B. G. 1993, \aap, 271, 391
\bibitem[Combes et al.(1990)]{com90} Combes, F., Debbasch, F., Friedli, D., \& Pfenniger, D. 1990,
\aap, 233, 82
\bibitem[Combes \& Gerin(1985)]{com85} Combes, F., \& Gerin, M. 1985, \aap, 150, 327
\bibitem[Combes \& Sanders(1981)]{com81} Combes, F., \& Sanders, R. H. 1981, \aap, 96, 164
\bibitem[Corsini et al.(2003)]{cor03} Corsini, E. M., Pizzella, A., Coccato, L., \& Bertola, F. 2003,
\aap, 408, 873
\bibitem[Courteau, de Jong \& Broeils(1996)]{cou96} Courteau, S., de Jong, R. S., \& Broeils, A. H.
1996, \apjl, 457, 73
\bibitem[Crenshaw, Kraemer \& Gabel(2003)]{cre03} Crenshaw, D. M., Kraemer, S. B.,
\& Gabel, J. R. 2003, \aj, 126, 1690
\bibitem[Debattista et al.(2004)]{deb04} Debattista, V. P., Carollo, C. M., Mayer, L., \& Moore, B.
2004, \apjl, 604, L93
\bibitem[de Grijs \& Peletier(1997)]{deg97} de Grijs, R., \& Peletier, R. F. 1997, A\&AL, 320, L21
\bibitem[Dehnen \& Binney(1998)]{deh98} Dehnen, W., \& Binney, J. J. 1998, \mnras, 298, 387
\bibitem[Delhaye (1965)]{del65} Delhaye, J. 1965, Galactic Structure, ed. A. Blaauw \& M. Schmidt
(Chicago: University of Chicago Press)
\bibitem[de Souza \& dos Anjos(1987)]{des87} de Souza, R. E., \& dos Anjos, S. 1987, \aaps, 70, 465
\bibitem[de Souza, Gadotti \& dos Anjos(2004)]{des04} de Souza, R. E., Gadotti, D. A., \& dos Anjos, S.
2004, \apjs, 153, 411
\bibitem[de Vaucouleurs et al.(1991)]{dev91} de Vaucouleurs, G., de Vaucouleurs, A., Corwin, H. G.,
Buta, R. J., Paturel, G., \& Fouque, P. 1991, Third Reference Catalog of Bright Galaxies (New York:
Springer-Verlag {\bf (RC3)}
\bibitem[Edvardsson et al.(1993)]{edv93} Edvardsson, B., Andersen, J., Gustafsson, B., Lambert, D. L.,
Nissen, P. E., \& Tomkin, J. 1993, \aap, 275, 101
\bibitem[Elmegreen \& Elmegreen(1985)]{elm85} Elmegreen, B. G., \& Elmegreen, D. M. 1985, \apj, 288, 438
\bibitem[Elmegreen, Elmegreen \& Hirst(2004)]{elm04} Elmegreen, B. G., Elmegreen, D. M., \& Hirst, A. C.
2004, \apj, 612, 191
\bibitem[Emsellem et al.(2001)]{ems01} Emsellem, E., Greusard, D., Combes, F., Friedli, D., Leon, S.,
P\'econtal, E., \& Wozniak, H. 2001, \aap, 368, 52
\bibitem[Fisher(1997)]{fis97} Fisher, D. 1997, \aj, 113, 950
\bibitem[Freeman(1970)]{fre70} Freeman, K. C. 1970, \apj, 160, 811
\bibitem[Freeman(1978)]{fre78} Freeman, K. C. 1978, IAU Symp., 77, 3
\bibitem[Friedli \& Benz(1993)]{fri93} Friedli, D., \& Benz, W. 1993, \aap, 268, 65
\bibitem[Friedli \& Benz(1995)]{fri95} Friedli, D., \& Benz, W. 1995, \aap, 301, 649
\bibitem[Gadotti \& de Souza(2003)]{gad03} Gadotti, D. A., \& de Souza, R. E. 2003, \apjl, 583, L75
\bibitem[Gadotti \& de Souza(2005)]{gad05} Gadotti, D. A., \& de Souza, R. E. 2005, \apj, submitted
{\bf (Paper II)}
\bibitem[Gadotti \& dos Anjos(2001)]{gad01} Gadotti, D. A., \& dos Anjos, S. 2001, \aj, 122, 1298
\bibitem[Gerssen, Kuijken \& Merrifield(1997)]{ger97} Gerssen, J., Kuijken, K., \& Merrifield,
M. R. 1997, \mnras, 288, 618
\bibitem[Gerssen, Kuijken, \& Merrifield(2000)]{ger00} Gerssen, J., Kuijken, K., \& Merrifield,
M. R. 2000, \mnras, 317, 545
\bibitem[Goudfrooij \& Emsellem(1996)]{gou96} Goudfrooij, P., \& Emsellem, E. 1996, \aap, 306, L45
\bibitem[Ho, Filippenko \& Sargent(1997a)]{ho97a} Ho, L. C., Filippenko, A. V., \& Sargent,
W. L. W. 1997a, \apj, 487, 579
\bibitem[Ho, Filippenko \& Sargent(1997b)]{ho97b} Ho, L. C., Filippenko, A. V., \& Sargent,
W. L. W. 1997b, \apj, 487, 591
\bibitem[Hoffleit \& Warren(1991)]{hof91} Hoffleit D., \& Warren W. H. 1991, The Bright Star Catalogue
\bibitem[Jogee et al.(2004)]{jog04} Jogee, S. et al. 2004, \apjl, submitted (astro-ph/0408382)
\bibitem[Jungwiert, Combes \& Axon(1997)]{jun97} Jungwiert, B., Combes, F., \& Axon, D. J. 1997,
\aaps, 125, 479
\bibitem[Kalnajs(1972)]{kal72} Kalnajs, A. J. 1972, \apj, 175, 63
\bibitem[Kennicutt(1992)]{ken92} Kennicutt, R. C. 1992, \apjs, 79, 255
\bibitem[Knapen, Shlosman \& Peletier(2000)]{kna00} Knapen, J. H., Shlosman, I., \& Peletier, R. F.
2000, \apj, 529, 93
\bibitem[Kormendy(1982)]{kor82} Kormendy, J. 1982, \apj, 257, 75
\bibitem[Kormendy \& Illingworth(1983)]{kor83} Kormendy, J., \& Illingworth, G. 1983, \apj, 265, 632
\bibitem[Kormendy \& Kennicutt(2004)]{kor04} Kormendy, J., \& Kennicutt, R. C. 2004, \araa, in press
\bibitem[Kregel, van der Kruit \& Freeman(2004)]{kre04} Kregel, M., van der Kruit, P. C., \& Freeman,
K. C. 2004, \mnras, 351, 1247
\bibitem[Kuijken \& Merrifield(1995)]{kui95} Kuijken, K., \& Merrifield, M. R. 1995, \apjl, 443, L13
\bibitem[Laine et al.(2002)]{lai02} Laine, S., Shlosman, I., Knapen, J. H., \& Peletier, R. F. 2002,
\apj, 567, 97
\bibitem[Laurikainen, Salo \& Buta(2004)]{lau04} Laurikainen, E., Salo, H., \& Buta, R. 2004, \apj, 607, 103
\bibitem[Maraston(1998)]{mar98} Maraston, C. 1998, \mnras, 300, 872
\bibitem[Martin \& Roy(1994)]{mar94} Martin, P., \& Roy, J. R. 1994, \apj, 424, 599
\bibitem[Massey(1997)]{mas97} Massey, P. 1997, A User's Guide to CCD Reductions with IRAF
\bibitem[Massey, Valdes \& Barnes(1992)]{mas92} Massey, P., Valdes, F., \& Barnes, J.
1992, A User's Guide to Reducing Slit Spectra with IRAF
\bibitem[McElroy(1995)]{mce95} McElroy, D. B. 1995, \apjs, 100, 105
\bibitem[Merrifield(2004)]{mer04} Merrifield, M. R. 2004, ASP Conf. S., 317
\bibitem[Merrifield \& Kuijken(1999)]{mer99} Merrifield, M. R., \& Kuijken, K. 1999, \aap, 345, 47
\bibitem[Merritt(1996)]{mer96} Merritt, D. 1996, \aj, 111, 2462
\bibitem[Merritt \& Sellwood(1994)]{mer94} Merritt, D., \& Sellwood, J. A. 1994, \apj, 425, 551
\bibitem[Mulchaey \& Regan(1997)]{mul97} Mulchaey, J. S., \& Regan, M. W. 1997, \apj, 482, L135
\bibitem[Noguchi(2000)]{nog00} Noguchi, M. 2000, \mnras, 312, 194
\bibitem[Norman, Sellwood \& Hasan(1996)]{nor96} Norman, C. A., Sellwood, J. A., \& Hasan, H. 1996,
\apj, 462, 114
\bibitem[Oke(1990)]{oke90} Oke, J. B. 1990, \aj, 99, 1621
\bibitem[Patsis \& Athanassoula(2000)]{pat00} Patsis, P. A., \& Athanassoula, E. 2000, \aap, 358, 45
\bibitem[Peletier \& Balcells(1996)]{pel96} Peletier, R. F., \& Balcells, M. 1996, \aj, 111, 2238
\bibitem[Pfenniger \& Norman(1990)]{pfe90} Pfenniger, D., \& Norman, C. 1990, \apj, 363, 391
\bibitem[Pizzella et al.(2004)]{piz04} Pizzella, A., Corsini, E. M., Vega Beltr\'an, J. C., \& Bertola, F.
2004, \aap, 424, 447
\bibitem[Polyachenko \& Polyachenko(2003)]{pol03} Polyachenko, V. L., \& Polyachenko, E. V. 2003,
AL, 29, 447
\bibitem[Prugniel \& Soubiran(2001)]{pru01} Prugniel, Ph., \& Soubiran, C. 2001, \aap, 369, 1048
\bibitem[Rix \& White(1992)]{rix92} Rix, H-W., \& White, S. D. M. 1992, \mnras, 254, 389
\bibitem[Rubin et al.(1985)]{rub85} Rubin, V. C., Burstein, D., Ford, W. K., \& Thonnard, N. 1985, \apj,
289, 81
\bibitem[Sakamoto et al.(1999a)]{sak99a} Sakamoto, K., Okumura, S. K., Ishizuki, S., \& Scoville, N. Z. 1999a,
\apj, 525, 691
\bibitem[Sakamoto et al.(1999b)]{sak99b} Sakamoto, K., Okumura, S. K., Ishizuki, S., \& Scoville, N. Z. 1999b,
\apjs, 124, 403
\bibitem[Sandage \& Bedke(1994)]{san94} Sandage, A., \& Bedke, J. 1994, The Carnegie Atlas of
Galaxies, Vol. 1 (Washington, DC: Carnegie Inst.)
\bibitem[Schwarz(1981)]{sch81} Schwarz, M. P. 1981, \apj, 247, 77
\bibitem[Sellwood \& Moore(1999)]{sel99} Sellwood, J. A., \& Moore, E. M. 1999, \apj, 510, 125
\bibitem[Sellwood \& Wilkinson(1993)]{sel93} Sellwood, J. A., \& Wilkinson, A. 1993,
Rep. Prog. Ph., 56, 173
\bibitem[S\'ersic \& Pastoriza(1965)]{ser65} S\'ersic, J. L., \& Pastoriza, M. 1965, \pasp, 77, 287
\bibitem[S\'ersic \& Pastoriza(1967)]{ser67} S\'ersic, J. L., \& Pastoriza, M. 1967, \pasp, 79, 152
\bibitem[Shen \& Sellwood(2004)]{she04} Shen, J., \& Sellwood, J. A. 2004, \apj, 604, 614
\bibitem[Sheth et al.(2003)]{she03} Sheth, K., Regan, M. W., Scoville, N. Z., \& Strubbe, L. E. 2003,
\apjl, 592, L13
\bibitem[Shlosman, Begelman \& Frank(1990)]{shl90} Shlosman, I., Begelman, M. C., \& Frank, J. 1990,
\nat, 345, 679
\bibitem[Shlosman, Frank \& Begelman(1989)]{shl89} Shlosman, I., Frank, J., \& Begelman, M. C. 1989,
\nat, 338, 45
\bibitem[Spitzer \& Schwarzschild(1951)]{spi51} Spitzer, L., \& Schwarzschild, M. 1951, \apj, 114, 385
\bibitem[Spitzer \& Schwarzschild(1953)]{spi53} Spitzer, L., \& Schwarzschild, M. 1953, \apj, 118, 106
\bibitem[Teuben(1995)]{teu95} Teuben, P. J. 1995, ASP Conf. Ser., 77, 398
\bibitem[Tinsley \& Gunn(1976)]{tin76} Tinsley, B. M., \& Gunn, J. E. 1976, \apj, 203, 52
\bibitem[Tonry \& Davis(1979)]{ton79} Tonry, J., \& Davis, M. 1979, \aj, 84, 1511
\bibitem[Toomre(1963)]{too63} Toomre, A. 1963, \apj, 138, 385
\bibitem[Toomre(1964)]{too64} Toomre, A. 1964, \apj, 139, 1217
\bibitem[Toomre(1966)]{too66} Toomre, A. 1966, Notes on the 1966 Summer Study Program in
Geophysical Fluid Dynamics at Woods Hole Oceanographic Institution, Vol. 1 (Woods Hole:
Woods Hole Oceanographic Inst.), 111
\bibitem[Toomre(1981)]{too81} Toomre, A. 1981, Structure and Evolution of Normal
Galaxies, ed. S. M. Fall \& D. Lynden-Bell (Cambridge: Cambridge Univ. Press), 111
\bibitem[van den Bergh(2002)]{vdb02} van den Bergh, S. 2002, \aj, 124, 782
\bibitem[van der Kruit(2002)]{vdk02} van der Kruit, P. C. 2002, ASP Conf. S., 273, 7
\bibitem[van der Kruit \& Searle(1981)]{vdk81} van der Kruit, P. C., \& Searle, L. 1981, \aap, 95, 105
\bibitem[van der Marel \& Franx(1993)]{vdm93} van der Marel, R. P., \& Franx, M. 1993, \apj, 407, 525
\bibitem[Villumsen(1983)]{vil83} Villumsen, J. V. 1983, \apj, 274, 632
\bibitem[Wielen(1977)]{wie77} Wielen, R. 1977, \aap, 60, 263
\bibitem[Zaritsky, Kennicutt \& Huchra(1994)]{zar94} Zaritsky, D., Kennicutt, R. C., \& Huchra, J. P. 1994,
\apj, 420, 87
\end{thebibliography}
\end{document}